\documentclass[sigconf]{acmart} 
\AtBeginDocument{
  }
\copyrightyear{2025}
\acmYear{2025}
\setcopyright{cc}
\setcctype{by}
\copyrightyear{2025}
\acmYear{2025}
\setcopyright{cc}
\setcctype{by}
\acmConference[CHI '25]{CHI Conference on Human Factors in Computing Systems}{April 26-May 1, 2025}{Yokohama, Japan}
\acmBooktitle{CHI Conference on Human Factors in Computing Systems (CHI '25), April 26-May 1, 2025, Yokohama, Japan}\acmDOI{10.1145/3706598.3714090}
\acmISBN{979-8-4007-1394-1/25/04}

\usepackage{enumitem}
\usepackage{longtable}
\usepackage{booktabs}
\usepackage{tabularx}

\AtBeginDocument{
  }

\acmConference[CHI '25]{The ACM CHI Conference on Human Factors in Computing Systems}{April 26--May 1, 2025}{Yokohama, Japan}

\begin{document}

\title[Motivations, Practices, Challenges, and Opportunities for Cross-Functional Collaboration around AI within the News Industry]{"It Might be Technically Impressive, But It’s Practically Useless to us": Motivations, Practices, Challenges, and Opportunities for Cross-Functional Collaboration around AI within the News Industry}

\author{Qing Xiao}
\affiliation{
  \institution{Human-Computer Interaction Institute, Carnegie Mellon University}
  \city{Pittsburgh}
  \state{Pennsylvania}
  \country{USA}
}
\email{qingx@andrew.cmu.edu}

\author{Xianzhe Fan}
\affiliation{
  \institution{Tsinghua University}
  \city{Beijing}
  \country{China}
}
\email{fxz21@mails.tsinghua.edu.cn}

\author{Felix M. Simon}
\affiliation{
  \institution{Reuters Institute for the Study of Journalism, University of Oxford}
  \city{Oxford}
  \country{United Kingdom}
}
\email{felix.simon@politics.ox.ac.uk}

\author{Bingbing Zhang}
\affiliation{
  \institution{School of Journalism and Mass Communication, University of Iowa}
  \city{Iowa City}
  \state{Iowa}
  \country{USA}
}
\email{bingbing-zhang@uiowa.edu}

\author{Motahhare Eslami}
\affiliation{
  \institution{Human-Computer Interaction Institute, Carnegie Mellon University}
  \city{Pittsburgh}
  \state{Pennsylvania}
  \country{USA}
}
\email{meslami@andrew.cmu.edu}

\begin{abstract}
Recently, an increasing number of news organizations have integrated artificial intelligence (AI) into their workflows, leading to a further influx of AI technologists and data workers into the news industry. This has initiated cross-functional collaborations between these professionals and journalists. Although prior research has explored the impact of AI-related roles entering the news industry, there is a lack of studies on how internal cross-functional collaboration around AI unfolds between AI professionals and journalists within the news industry. Through interviews with 17 journalists, six AI technologists, and three AI workers with cross-functional experience from leading Chinese news organizations, we investigate the practices, challenges, and opportunities for internal cross-functional collaboration around AI in news industry. We first study how these journalists and AI professionals perceive existing internal cross-collaboration strategies. We explore the challenges of cross-functional collaboration and provide recommendations for enhancing future cross-functional collaboration around AI in the news industry.
\end{abstract}


\maketitle

\newcommand{\add}[1]{\textcolor{black}{#1}}
\newcommand{\newadd}[1]{\textcolor{black}{#1}}

\section{Introduction}

Recently, an increasing number of news organizations have incorporated Artificial Intelligence (AI) into journalistic workflows. A global survey for 2023 by the London School of Economics and Political Science revealed that more than 75\% of the news organizations sampled in 46 countries use AI somewhere in their work processes and about a third actively developing or having AI strategies at the time~\cite{journalismai_ai_2023}. Previous studies have explored the collaboration of newsrooms with external technology companies to develop AI systems, demonstrating that such partnerships undermine the independence of the news industry~\cite{simon2024escape, journalismai_ai_2023, simon_ai_2024, simon2022uneasy}. As a result, many newsrooms have shifted toward building internal AI systems to integrate with their journalism practices~\cite{shlab2023, deAssis2024AI, Thrasher2024, Bloomberg2023}. This internal integration has led to the creation of new roles within news organizations, such as AI experts, data scientists, and other specialized data workers, reshaping the structure and workflow of the industry.

Although this internal integration of AI into newsrooms has created numerous opportunities for innovation and efficiency~\cite{parasie2022computing}, it has also introduced new challenges, such as a lack of mutual expertise - where journalists have limited knowledge of AI and AI professionals have limited understanding of journalism, goal misalignments between journalists and AI experts, and power imbalances. For example, the application of AI in journalism is consistently based on technical experts with advanced AI knowledge who can build or integrate AI systems and AI workers who handle data annotation tasks~\cite{de2021artificial, gutierrez2023question, stenbom2023exploring}. However, these individuals often have limited knowledge of the practice, routines, and norms of journalism and reporting work. Journalists, meanwhile, can have limited technical expertise. Although more up-to-date data is scarce, a 2017 survey by The International Center for Journalists (ICFJ), which surveyed over 2,700 newsroom employees from 130 countries, found that only 5\% of newsroom employees had degrees in technology-related fields such as computer science and only 9\% of technologists working in newsrooms have a journalism or communications degree~\cite{ICFJ2017}. The integration of AI has also revealed a complex dynamic in which the capabilities and limitations of AI interact with and sometimes clash with traditional newsroom norms~\cite{simon2023policies, gutierrez2023question, lee2020predicting, stenbom2023exploring}. AI technologists and more technical workers entering the news industry may lack a deeper understanding of the routines and needs of journalists, leading them to develop technological systems and tools that do not fully align with the needs of journalists, potentially disrupting established workflows and practices or developing systems that are of little use and abandoned quickly~\cite{simon2024escape}. Power dynamics, where journalists and editors tend to hold the primary gatekeeping role in the process of technology adoption, can become another obstacle, sometimes marginalizing technical personnel in newsrooms~\cite{min2021keeping, aronson2015virtual, gutierrez2023question,simon2024escape}. 

Given these challenges in the ongoing integration of AI within the news industry, HCI scholars have underscored the urgent need for effective cross-functional collaboration between journalists and AI technologists~\cite{cai2021onboarding, aitamurto2019hci}. Such collaboration is crucial for developing AI-driven news systems that effectively integrate ethical standards and journalistic principles into systems meant to benefit the work of news organizations~\cite{nishal2024blueprints}. Previous studies have shown that AI integration in various industries often necessitates extensive collaboration between different roles, yet this process can be hindered by differing perceptions, expectations, and goals of various roles~\cite{yang2020re, deng2023investigating, almahmoud2021teams, deng2023understanding}. Despite these insights, it remains unclear how AI integration has specifically impacted cross-functional collaboration within newsrooms and how it has reshaped interactions and workflows across different departments and roles in the news industry. 

\add{Prior research has partially addressed internal cross-functional collaboration within news organizations, but key gaps remain. For instance, Lopez et al. examined the integration of AI into BBC's workflows, highlighting the need for alignment between journalists and technologists to meet newsroom objectives~\cite{gutierrez2023question}. Moller explored the dominance of developers and data scientists in algorithmic curation due to limited collaboration with journalists, finding that the development of recommendation systems often occurs in isolation~\cite{moller2023bridging}. Additionally, Lewis and Usher analyzed the distinct perspectives of programmers and journalists, showing how boundary negotiations shaped their collaboration~\cite{lewis2016trading}.} \newadd{While these studies underscore the significance of cross-functional collaboration, they focus on specific roles or isolated processes rather than providing a systematic analysis about how different roles collaborate with each others, especially around AI.}

Therefore, our study raises the following research questions:
\add{\begin{itemize}
 \item \textbf{RQ1:} What are the (a) motivations and (b) current practices behind internal cross-functional collaboration around AI within newsrooms?
 \item \textbf{RQ2:} (a) What challenges do such internal cross-functional collaborations around AI face, and (b) what potential solutions and opportunities exist to address these challenges?
\end{itemize}}

To explore these issues, we conducted a two-stage study with 26 professionals from five leading Chinese news organizations, including 17 journalists, 6 AI technologists, and 3 AI workers (data annotators). First, we conducted semi-structured interviews to identify current collaboration strategies and challenges. Second, we held three workshops where participants collaboratively envisioned a more effective and equitable AI collaboration. We focused on Chinese news organizations due to their integration of LLM researchers, algorithm engineers, data scientists, and journalists~\cite{RenminRibao2023,PeoplePostdoc2020,GuangzhouDaily2023}, extensive use of AIGC~\cite{shu2023}, and in-house development of AI systems and GPTs~\cite{shlab2023}.

Our findings reveal that while outsourcing to tech companies works for specific projects, it limits newsroom autonomy, a concern as AI increasingly influences journalism. To counter this, organizations are recruiting AI technologists and promoting collaboration through lectures, workshops, and informal knowledge-sharing initiatives. However, co-design processes remain absent: journalists are excluded from AI tool design, and technologists struggle to assess their needs (\textit{\textbf{RQ1}}). In the workshops, participants emphasized the need for effective tools that align with journalistic workflows without disrupting them. They proposed strategies for improving co-design processes and highlighted the marginalized role of AI workers, who are excluded from meaningful collaboration and limited to data-related tasks. To foster equitable knowledge-sharing, we advocate for a more inclusive approach to cross-functional collaboration in the news industry (\textit{\textbf{RQ2}}).

\section{Related Work}
In this section, we first review studies on AI and journalism. Then, we discuss cross-functional collaboration in news organizations. Next, we review past research on cross-functional collaboration around AI in the HCI area. 

\subsection{AI in News Industry}
Recent research highlights AI's impact on the news industry ~\cite{aitamurto2019hci, simon2023policies, stenbom2023exploring, gutierrez2023question}. The advent of OpenAI's chatbot ChatGPT and related models by other companies has spurred interest in large language models (LLMs) and foundation models within many newsrooms ~\cite{simon2023policies, journalismai_ai_2023, Fletcher2024, pinto2024artificial}, with many international news organizations adopting LLMs ~\cite{simon2023policies, bbc_ai_guidance_2024, Fletcher2024, pinto2024artificial}. In the opinion of some journalists and scholars, these AI technologies can help improve news production and distribution, the interaction with audiences, and the business models of news organizations~\cite{caswell2023ai, lee2020predicting, li2023newsdialogues}.

In the news industry, AI technologies such as computer vision (CV) and natural language processing (NLP) are employed across different stages of the news production and distribution process~\cite{simon2022uneasy,seychell2024ai,aljalabneh2024balancing, stenbom2023exploring}. \add{Drawn from gatekeeping theory in media and journalism studies, Simon~\cite{simon2022uneasy} identifies several key use cases where AI is utilized in the news industry throughout the gatekeeping process~\label{model}:}

\begin{itemize}
\item \add{\textbf{Information Selection:} AI is used to identify (1) relevant content by mining data from, e.g., social media, (2) detecting news stories, and (3) verifying information through image or video recognition systems. It also organizes data by categorizing, tagging, and generating metadata.}
\item \add{\textbf{News Production:} AI helps to automate news production tasks, such as generating stories, editing, formatting, translation, and transcription, or handling archival research.}
\item \add{\textbf{Content Distribution:} AI enables personalized and customized forms of news distribution through the use of, e.g., recommendation systems or the generation of tailored content, helps with tasks like website management, or enables new forms of audience interactivity such as Chatbots text-to-speech features.}
\item \add{\textbf{Audience and Business Analytics:} AI helps to analyze audience behavior to optimize advertising, tailor content, manage dynamic paywalls, and provide editorial insights.}
\end{itemize}

The increasing use of AI in the news industry has raised several concerns. First, journalists worry that AI systems may diverge from core journalistic values like accuracy, fairness, and public interest, which are crucial for maintaining credibility \cite{simon_ai_2024, simon2022uneasy}. Journalists often cannot adjust proprietary AI systems, making it difficult to align these tools with journalistic standards and potentially undermine news organizations' traditional gatekeeping roles\cite{simon_ai_2024, simon2022uneasy}. Other studies have shown that AI can amplify biases present in the training data, resulting in biased news coverage~\cite{leiser2022bias,broussard2019artificial, fang2024bias}. Second, the widespread adoption of AI, coupled with limited technological capabilities in newsrooms, could lead to forms of technological "media capture," where technology companies gain greater control over parts of the news production process, limiting newsrooms' autonomy~\cite{simon_ai_2024, simon2022uneasy, trattner2022responsible}. Additionally, AI use raises issues around the privacy and security of data, questions of intellectual property, and public trust in news~\cite{Fletcher2024,kothari2022artificial, thurman2017reporters, wolker2021algorithms}.

To reduce the harms associated with the deployment of AI, news organizations have implemented several strategies. Many have adopted ethical guidelines emphasizing transparency, accountability, and fairness to ensure that AI tools align with journalistic values~\cite{publicmedia2024, bbc_ai_guidance_2024, simon2023policies}. In some newsrooms, journalists are also receiving training to better understand AI technologies, their limitations, and ethical implications~\cite{jones2024karen}. Other organizations are partnering with technology companies or academic institutions to achieve this goal~\cite{csu2024generativeAI}. 

Finally, in an effort to build AI tools and systems better aligned with their unique needs and journalistic values and reduce their dependence on external technology companies, some news organizations are developing their own AI systems~\cite{simon2024escape, simon_ai_2024, publicmedia2024, Kahn2024, Bloomberg2023,Thrasher2024,deAssis2024AI}. This involves hiring skilled AI technologists and AI workers who are specifically tasked with designing, developing, maintaining, and optimizing AI systems that meet the unique needs of the newsrooms~\cite{gutierrez2023question}.

In all these examples, teams and individuals from different disciplinary backgrounds and with different (technical skills) must engage in complex processes of collaboration, with AI technologists having to design and develop AI products and tools tailored for journalistic tasks and journalists having to decide what they require and whether and how to use these tools. According to extant HCI research, cross-functional collaborations between AI practitioners and domain experts can offer a promising remedy for bridging the gap between technological capabilities and specific domain needs (e.g. in responsible AI practices) ~\cite{aitamurto2019hci,deng2023understanding,deng2023investigating}. However, current research has yet to explore how cross-functional collaboration around AI is actually taking place within the news industry and the specific challenges these collaborations face.

\newadd{While extensive research has explored AI's role in the news industry, including its applications, ethical implications, and organizational responses, there remains a significant gap in understanding the internal cross-functional collaborations that drive these advancements about AI. Specifically, how journalists and AI technologists navigate their differing expertise, priorities, and challenges to co-develop AI tools for journalistic tasks is understudied. Addressing this gap is essential, as these collaborations directly influence the integration of AI into newsroom workflows and ensure alignment with journalistic values.}

\subsection{Cross-functional Collaboration in News Organizations}
 
Past research has shown that journalism tends to embrace emerging technologies selectively, adopting new tools when they can clearly enhance the core journalistic values~\cite{dodds2024impact,akpe2024immersive, aronson2015virtual}. This conditional acceptance is often accompanied by close collaboration between journalists and technologists~\cite{akpe2024immersive}.  Journalists are responsible for overseeing and evaluating the effectiveness of the technology, communicating their ideas and requirements to the technical team. Meanwhile, technologists are tasked with explaining the technical possibilities of developing and implementing these tools.

For instance, Aronson~\cite{aronson2015virtual} has shown that the development of Virtual Reality (VR) news content requires a collaborative effort between journalists and VR technologists. While the former needs to master the unique narrative structures and affordances of VR storytelling, VR technologists play a critical role in managing the technical complexities involved in VR news production, including more intricate post-production processes. They also need to address the technical challenges associated with creating a VR experience that is both engaging and technically feasible, including considerations of hardware limitations, user interface design, and optimizing for different VR platforms~\cite{aronson2015virtual, doyle2016viewing}. Only through a cross-functional collaborative approach can both the journalistic integrity of the content and the technical quality of the VR experience be effectively maintained, in the service of creating compelling and innovative news stories.

Previous HCI studies have already explored how to improve journalists' AI literacy, such as through AI toolkits ~\cite{cai2023motivations} or the creation of human-centered tools to assist them in their work ~\cite{maiden2019evaluating,nishal2024understanding}. However, these studies have limitations. As Yang et al. points out, practitioners should mainly understand the capabilities and limitations of the technology and envision its future possibilities, while engineers are responsible for the technical feasibility and capabilities ~\cite{yang2020re}. Simply improving the technical literacy of domain experts (e.g., journalists' AI literacy in journalism) and making them aware of too many technical details is not the key to collaborative design. Instead, it can add unnecessary burdens and provide limited effectiveness~\cite{yang2020re}. 

\add{Additionally, while not originally developed in the context of AI, the concept of trading zones~\cite{collins2007trading,shrager2010wizards,dodds2024collaborative} offers a valuable theoretical lens for understanding the cross-functional collaborations that occur as news organizations adopt and adapt new technologies. Trading zones, defined as spaces where groups with different expertise, goals, and values negotiate shared frameworks of understanding and practice, have been used to explain how newsrooms integrate emerging professional practices and tools into their existing workflows~\cite{collins2007trading,kasavin2017trading}. Within these zones, boundary objects, interactional expertise, and intermediate languages help journalists, technologists, and other stakeholders collaboratively develop solutions that maintain journalistic integrity while accommodating technological constraints~\cite{lewis2016trading,dodds2024collaborative}.}

~\add{A salient example of this dynamic is found in Lewis and Usher’s examination of the Knight-Mozilla News Technology partnership~\cite{lewis2016trading}. They~\cite{lewis2016trading} focus on a key stage of an online 'Learning Lab', where journalists and programmers converged around a shared interest in advancing journalism through open source software. In doing so, they illustrate how “distinct understandings about~\cite{lewis2016trading} news and technology converged, diverged, and ultimately blended” as participants worked toward journalism models that were “more process-oriented, participatory and socially curated~\cite{lewis2016trading}.” This case underscores how boundary negotiations play out in news practice as programmers, bringing their ethics and values, enter the journalistic field, influencing the norms and values at the core of the leading news institutions~\cite{lewis2016trading}.}

\newadd{Our study contributes to this growing body of research by applying the concept of trading zones~\cite{collins2007trading,zamith2019algorithms} and cross-functional collaboration~\cite{zhang2020data} in the integration of AI technologies in the newsroom. Rather than focusing solely on journalist technical literacy or engineer understanding of newsroom workflows, we investigate how cross-functional teams collaboratively negotiate the practice necessary for meaningful AI integration. We build on the framework of trading zones~\cite{collins2007trading,zamith2019algorithms} to illustrate how journalists and technologists collaboratively bridge epistemic and operational gaps, resulting in more effective, just, and innovative AI-supported journalism.}

\subsection{Cross-functional Collaboration Around AI}

In HCI research, various tools and systems have been developed to promote cross-functional collaboration ~\cite{zhang2020data}. These tools, such as task tracking, brainstorming and collaborative design applications, are based on cross-functional thinking and aim to provide integrated platforms for communication, collaboration, and knowledge management ~\cite{tahaei2021privacy}. However, when applied to AI practices, these cross-functional collaboration approaches often fall short.

Deng et al. examined the reasons for these failures and identified key issues, such as differing understanding of AI in various roles and inconsistent expectations between team members ~\cite{deng2023understanding, deng2023investigating}. There is a widespread difficulty among designers and users in understanding the capabilities and limitations of AI, which hinders brainstorming sessions and the initial stages of design ~\cite{dove2017ux, yang2018mapping, yang2018investigating, yang2018grounding}. Secondly, AI experts often struggle to grasp the needs of users and the contexts in which AI applications are deployed ~\cite{dove2017ux}. Furthermore, AI-specific prototypes frequently overlook the actual impact on end users, complicating the process of iterative improvement ~\cite{stone2016embedding}. Yang et al. further highlighted that in human-AI interaction, designers face challenges in creating AI systems due to unexpected reasoning errors and ethical dilemmas ~\cite{yang2020re} and the design of AI systems can lead to unforeseen ethical, fairness and other moral issues ~\cite{yang2020re}. These challenges affect not only AI experts but also collaborators and users from non-AI fields.

Furthermore, professionals outside the AI field still face difficulties in collaborating effectively with AI engineers. These challenges are often due to a lack of shared workflows, boundary objects, such as artifacts, documents, or concepts that serve as mutual reference points, and non-technical languages that facilitate collaboration ~\cite{yang2018mapping, yang2018investigating, yang2018grounding, yang2019sketching, yang2020re}. The absence of these elements creates a significant barrier to effective cross-functional collaboration in AI contexts.

\newadd{While existing HCI literature has highlighted general issues in human-AI interaction, we explore the newsroom as a distinct sociotechnical environment characterized by high-stakes public communication, and the integration of journalistic norms with technological constraints. Unlike other domains where AI adoption primarily optimizes efficiency or automation, newsrooms require interdisciplinary teams to navigate competing priorities, such as editorial independence, accuracy, and algorithmic optimization, while adapting to audience expectations. By addressing these context-specific challenges, our study contributes to a broader understanding of how to design AI systems that are not only technically feasible but also ethically and practically aligned with domain-specific needs, providing actionable insights for HCI researchers and practitioners.}

\section{Background: AI in the Chinese News Industry}

\add{The Chinese news industry and the innovation environment in China, where numerous experimental AI technologies have been seamlessly integrated into formal journalistic workflows ~\cite{yu2017}, presents an excellent case for exploring internal cross-functional collaborations centered around AI.} This background section provides context on both to better situate our findings and their relevance.

Various Chinese news organizations have made considerable progress in adopting AI over the past decade. For example, Xinhua News Agency, China's largest and official state news agency, introduced fully automated AI writing robots in 2015, applying them to collect real-time data and match news templates for limited automated content writing ~\cite{zhong2019}. In 2017, the "Media Brain" system, developed by Xinhua Zhiyun, was launched, marking the beginning of AI-assisted work in clue discovery, editorial production, effect feedback, and public opinion monitoring ~\cite{shen2018, liu2018}. In 2018, Xinhua News Agency also introduced the world’s first AI news anchor, designed to deliver news naturally and in a human-like way and improve its performance by learning from videos ~\cite{kothari2022artificial}. Since 2019, the application of LLMs in Chinese journalism has gained more importance, with such tools now widely used to quickly summarize information from policy and financial reports and extract key information from audio and video in specialized news production processes ~\cite{zhong2019, zhao2021}. In July 2023, China Media Group (CMG) launched CMG Media GPT, which leverages the extensive database of Chinese official media to generate news text, images, videos, and even digital human anchor news scenarios to augment news production at CMG ~\cite{shlab2023}.

\add{This widespread integration of AI into journalistic workflows is deeply intertwined with the Chinese government's broader use of AI as a strategic tool for governance, economic growth, and ideological influence~\cite{ding2018deciphering}. Chinese news organizations not only leverage AI technologies for operational efficiencies but also align their adoption with state priorities. The dual role of AI in supporting journalistic innovation and state governance reflects the CCP’s broader vision of harnessing emerging technologies to reinforce its socio-political and economic objectives~\cite{ding2018deciphering,zeng2020artificial}.}

AI plays a significant role in supporting the governance model of the Chinese Communist Party (CCP) and the state. First, AI is used to improve the state's control over information and surveillance, aligning with the CCP's focus on maintaining social stability and strengthening its ideological influence~\cite{zeng2020artificial}. Second, AI is integral to China’s long-term ambition to boost its economic and technological dominance. The CCP views AI as a "frontier industry," central to achieving its goal of becoming the global leader in AI technology by 2030~\cite{kuo2024china,ding2018deciphering}. Finally, AI supports the Chinese state’s emphasis on information management, particularly in the media. AI-generated content is already extensively used in state-controlled media to produce news that aligns with government narratives and strengthens the CCP’s control over public discourse~\cite{zeng2020artificial}.

Consequently, the regulatory landscape in China has also evolved alongside these technological advancements. Since January 1, 2020, China has criminalized the creation of misleading content using AI to reduce unethical uses of this technology~\cite{kothari2022artificial}. In 2023, the Chinese government issued additional regulations on AI in news dissemination, further encouraging the extensive application of generative AI in news production ~\cite{cac2023}. To support these advancements, many Chinese news organizations are continuously recruiting AI professionals such as LLM researchers and data scientists~\cite{RenminRibao2023,PeoplePostdoc2020,GuangzhouDaily2023}. AIGC has already become a routine part of Chinese news production ~\cite{peopledaily2023, tencent_ai_news_2024}. In March 2024, Chinese state television media released nearly 20 news stories using AIGC footage, averaging one to two AIGC reports every day throughout the month ~\cite{shu2023}. 

\section{Method}

To investigate how journalists, AI technologists and AI workers collaborate cross-functionally in China's current news industry, and to understand the challenges they face and ways to improve such collaborations, we conducted a two-phase study involving 17 journalists (J1-J17), six AI technologists (T1-T6) and three AI workers (W1-W3, data annotators in this study) from five leading Chinese news organizations. All participants had more than 12 months of experience in AI-driven news production and cross-functional collaboration within their respective organizations. The first phase of the study consisted of semi-structured one-on-one interviews, each lasting approximately 90 minutes. In the second phase, we organized three workshops where journalists, AI technologists, and AI workers discussed strategies to enhance cross-functional collaborations around AI in the news industry. This study received approval from the Institutional Review Board (IRB). In the following, we provide an overview of the participants and recruitment strategy, the design and conduct of the research interviews and workshops, as well as data analysis.

\add{To ensure compliance with IRB requirements and ethical standards, all participants are identified using anonymized IDs, and their ages and years of professional experience in the news industry have been grouped into ranges (e.g., 1-5, 6-10) to further obscure identifiable details. Additionally, the names of news organizations have been replaced with anonymized codes. These steps were implemented to minimize the risk of identifying individuals or their affiliations while maintaining the integrity of the research data.}

\subsection{Participants}

The deployment of AI in the Chinese news industry varies significantly between organizations and regions. Leading news institutions in China, such as People's Daily and CCTV, have established internal teams of AI technologists and supporting technology infrastructure. In contrast, some local newspapers lack technical personnel. Given this study's focus on cross-functional collaboration within the news industry, we employed purposive sampling to identify organizations and participants who likely would have had to collaborate across their organizations on AI. The primary criteria for selecting participants were: (1) individuals with at least one year of experience integrating AI technologies within their news organizations, (2) those actively involved in cross-functional collaboration between journalistic and AI teams in the past year, and (3) current members of leading Chinese news organizations known for their advanced AI applications and established collaboration mechanisms. Through personal connections and further snowball sampling \cite{noy2008sampling}, we ended up recruiting 17 journalists, six AI technologists, and three AI workers from five Chinese news organizations. Detailed demographic information can be found in Appendix (Table ~\ref{tab:FormativeParticipant}). 

Initially, our goal was not specifically to include AI workers (data annotators) in our study, as previous research and media reports rarely mentioned their contributions to AI in the news industry. However, during recruitment, several journalists and AI technologists highlighted the importance of these workers in the Chinese context. With their help, we recruited three AI workers, all of whom served as data annotators, a critical yet often overlooked role in training AI models. These annotators represent a broader group of support personnel, such as content moderators, whose contributions are essential, but frequently go unnoticed. By including them, we aim to highlight the experiences of AI workers in cross-functional collaboration within the news industry.

\add{During recruitment, the cross-functional collaboration experiences we aimed to capture included, but were not limited to, journalists working with AI technologists on designing automated content generation tools, AI workers preparing datasets for machine learning models for technologists, and journalists collaborating with developers to integrate recommendation systems into newsroom workflows. These experiences also included joint workshops, co-design sessions, or informal interactions where journalists and technologists shared knowledge to align AI tools with journalistic goals.}

\add{Moreover, despite the predominance of male AI technologists and younger AI workers compared to journalists, we prioritized diversity in our sample to reflect industry norms and capture varied perspectives. Although journalist participants were diverse in our study, the lack of female AI technologists in the organizations interviewed highlighted persistent gender inequality in STEM. In particular, all AI workers we interviewed were women, offering a contrasting view of gender representation. This imbalance underscores the uneven distribution of opportunities within the industry. Thus, workshop participants were also encouraged to reflect on gender and labor equity in AI-driven journalism.}

\subsection{Study Design}
\subsubsection{Interview}

We conducted one-on-one semi-structured interviews with all 26 participants, with each interview lasting approximately one and a half hours. The interviews were conducted remotely through video calls. 

\add{The interviews began with collecting actual examples and personal thoughts on cross-functional collaboration around AI in the news industry. Participants were asked to reflect on how these experiences relate to intrinsic values such as news value and AI fairness. We asked them to describe specific scenarios of cross-functional collaboration, focusing on their experiences at different stages of the collaborative process, such as initial planning, communication, and implementation. Participants were also encouraged to articulate their perspectives, reasoning, and specific strategies employed to navigate these collaborations. In addition, they shared insights on how their individual experiences and organizational cross-functional collaborations have shaped the practices of their newsrooms.}

\add{Our interview protocol was guided by the AI-driven journalism gatekeeping model outlined in \autoref{model}, which builds upon traditional gatekeeping theories while incorporating the key scenarios of AI application in modern journalism~\cite{simon_ai_2024, simon2022uneasy}. We used this specific gatekeeping model about AI usage in current newsrooms to frame our questions and guide the exploration of participants’ cross-functional collaboration experiences with various AI applications. For example, we consistently asked participants to reflect on specific instances where they encountered challenges or opportunities in AI-related tasks, examining the alignment between AI-driven processes and journalistic values.}

\add{In addition to the gatekeeping model, our interview design drew on methodologies from prior studies on journalism and cross-functional collaboration~\cite{deng2023understanding, deng2023investigating, nguyen2018cross, ambos2013combining}. This allowed us to investigate potential collaboration issues faced by participants, including role-specific tensions, communication barriers, and strategies to overcome such challenges. }

\add{The interview script was designed to address our two research questions directly. For RQ1, which explores the motivations and practices behind cross-functional collaboration, we included questions such as: \textit{"Can you describe a recent example of collaboration with colleagues from other roles (e.g., journalists, AI technologists) in your organization?"}} \add{For RQ2, which investigates the challenges of collaboration and potential solutions, we asked questions like: \textit{“What are the main challenges you have faced when working with colleagues from different roles or technical backgrounds?"}}

\subsubsection{Workshop}

\add{In the second stage, we conducted three 90-minute online workshops with five journalists, three AI technologists, and one AI worker. The same people participated in three sequential workshops (i.e., J3, J4, J13, J15, J17, T3, T4, T6, W3). We will refer to the different workshops as WS1, WS2, and WS3. Participants were encouraged to critically reflect on the present and envision future cross-functional collaboration centered on their experiences and values~\cite{holbert2020afrofuturism}. Inspired by Deng et al., we employed a human-centered design (HCD) approach~\cite{deng2023understanding, deng2023investigating} to structure the workshops. HCD emphasizes iterative participatory processes ~\cite{deng2023understanding, deng2023investigating} that center the perspectives and needs of the end users, making it particularly suitable for the promotion of collaboration between disciplines. Following this framework, we began each workshop by presenting participants with the cross-functional collaboration challenges identified in the first one-on-one interview phase. This HCD approach provided a foundation for structured reflection, allowing participants to build a shared understanding of these issues and collaboratively explore possible solutions.}

\add{The three workshops were designed to complement one another by progressively deepening the level of engagement and exploration. In the first workshop, participants were encouraged to devise practical strategies to address the challenges identified during the interview phase. This session served as a brainstorming exercise, focusing on immediate and actionable solutions to improve cross-functional collaboration.}

\add{The second workshop adopted a more disruptive approach with speculative design ideas, aiming to challenge participants to critically question existing practices and the underlying organizational structures in their newsrooms and the broader news industry. Participants were asked to reconsider established norms, explore systemic barriers, and identify areas where current collaborative practices failed. This session prioritized critical reflection over immediate solutions, fostering a deeper understanding of the root causes of collaboration challenges.}

\add{The third workshop shifted the focus toward envisioning the future. The participants were tasked with envisioning ideal collaborative environments and discussing long-term strategies to achieve these goals. This session emphasized aspirational thinking, encouraging participants to synthesize insights from previous workshops and articulate a vision for more collaborative and effective cross-functional practices.}

\add{In contrast to our one-on-one interview stage, which focused on personal experiences with collaboration and the challenges faced by individuals, the workshops emphasized collaborative exploration and collective reflection and imagination to address these difficulties. Unlike interviews centered on individual perspectives, workshops provided a dynamic space for interaction, where journalists, AI technologists, and AI workers could openly discuss and debate strategies to improve collaboration. Through these sessions, participants were able to collectively identify systemic issues, critique existing practices, and envision future scenarios for cross-functional collaboration.}

\subsection{Data Capture and Analysis}
\add{All interviews and workshops were conducted in Chinese to accommodate the participants' backgrounds. Two researchers then transcribed and translated the recordings into English. We used Atlas.ti for thematic analysis of both interview and workshop data, employing a combination of deductive and inductive approaches~\cite{fereday2006demonstrating}. This hybrid method allowed us to systematically examine the data based on predefined research questions while remaining open to emerging themes~\cite{fereday2006demonstrating}.}

\add{In line with qualitative research traditions, we chose not to calculate inter-coder reliability metrics, as our methodological approach prioritized depth, reflexivity, and contextual understanding over standardization~\cite{chinh2019ways,kuo2023understanding}. Qualitative research often embraces an interpretive paradigm, where the goal is not to quantify agreement but to uncover nuanced insights through an iterative engagement with the data~\cite{morgan2020iterative}. Thematic analysis in this study was conducted using an iterative process driven by the researchers' interpretations, allowing flexibility in identifying and refining themes as the analysis progresses~\cite{fereday2006demonstrating}. We ensured consistency and reliability through collaborative coding sessions. Two researchers worked together to analyze data segments, openly discussed their interpretations, and reached a consensus on emergent themes. This collaborative approach not only mitigated potential biases but also enriched the analysis by integrating diverse perspectives.}

\add{To further validate our findings, we employed a participant-researcher collaboration approach~\cite{goldstein2017reflexivity, boivin2018breaking, chase2017}. We shared initial coding results with participants, inviting them to confirm whether the themes accurately reflected their views and experiences. Based on their feedback, we refined the codes to enhance accuracy and credibility. This iterative engagement ensured that the final themes were grounded in participants' perspectives while aligning with the research objectives.}

\add{Additionally, our reflexive approach~\cite{chinh2019ways,kuo2023understanding} emphasized the researchers' critical examination of their positionality and influence on the analysis. Throughout the process, we documented coding decisions and reflexive notes to maintain transparency and ensure that the analysis remained grounded in the data rather than preconceived notions. This approach aligns with previous practices in qualitative research~\cite{kuo2023understanding}, emphasizing trustworthiness and credibility over statistical measures. In the following ~\autoref{Motivation}, \autoref{practices} and \autoref{challenges}, we outline the key themes that emerged from our data.}

~\section{~\add{Interview Findings: Motivations for Establishing Cross-functional AI teams in News Industry (RQ1-a)}}~\label{Motivation}

\add{We found that the complexity of AI technology, coupled with its increasingly comprehensive applications in journalism, which surpasses the capabilities of traditional journalist training or outsourcing to external technologists, has become a primary driver for news organizations to establish internal cross-functional AI teams, enabling them to effectively navigate and leverage this disruptive innovation.}

\subsection{~\add{Mastering AI in the News Industry: A Challenge for Journalists, but an Essential Adoption}}

\newadd{According to our participants, most newsrooms in China must actively adapt to and integrate into the wave of news innovation about AI. This innovation-driven approach is not merely a response to market competition but is also closely tied to the country's overall development strategy.} \newadd{As J2 stated, \textit{"In China, 'technology is the primary productive force, and innovation is the primary driving force' has become a national-level development slogan, especially in president Xi Jinping era. The Chinese government places great emphasis on technological innovation. This principle not only guides the formulation of national science and technology policies but also profoundly influences the development trajectories of various industries in China, including the news industry."}} \newadd{Therefore, for the Chinese news industry, AI-driven innovation is not optional but essential to meet national strategic demands, but the specific pathways for innovation are left for news professionals to explore.} \newadd{As J2 stated, \textit{"China places a strong emphasis on innovation, mandating organizations to pursue technological advancements. At the same time, the question of how to innovate is largely left to industry professionals to figure out on their own. There is a saying in China: ‘crossing the river by feeling the stones (mozheshitouguohe).’"}} \newadd{This phrase proposed by the Chinese government metaphorically describes a gradual, experimental approach to problem-solving, where individuals navigate uncertainty step by step rather than following a predefined roadmap. It reflects a strategy of cautious adaptation from the Chinese government, allowing practitioners to adjust their methods based on real-world challenges and evolving conditions.} \newadd{J2 added, \textit{"For us, this means that while we cannot afford to lag behind or ignore AI, the way we integrate AI into our workflow is something we must explore progressively, learning from experience and adapting to emerging technological and professional demands."}} 

\newadd{The Chinese news industry faces significant pressure to innovate with AI, yet its previous approaches to technological innovation have proven largely unreliable when it comes to AI. As a result, the industry has shifted toward forming internal cross-functional teams around AI. Specifically, when discussing the motivation for internal cross-functional collaboration around AI, most participants (n = 13), including 9 journalists and 4 AI technologists, highlighted the unique challenges of AI compared to other emerging technologies. These challenges include (1) the complexity of AI, which journalists themselves find difficult to master, requiring collaboration with others, and (2) the ineffectiveness of traditional outsourcing practices in this domain, necessitating the establishment of internal teams.}

J1 mentioned \textit{"Before the AI boom, we also introduced many emerging technologies, such as data visualization. However, during those times, we did not hire AI technologists on a large scale."} Tools like Tableau, commonly taught in leading journalism schools, made technologies like data visualization accessible to journalists. As a result, Chinese news organizations rarely hired technologists in areas like data visualization and virtual reality, and demand for cross-functional collaboration was minimal. J1, J2, and J3, all senior managers, shared similar views. As J1 explained, \textit{"We find it hard to believe that technologists truly understand how to produce news. The best approach in our previous view is for journalists themselves to learn these technologies and apply them directly to news production."}

Even when technical support was needed, newsrooms often outsourced projects, forming temporary cross-functional teams. J2 elaborated, \begin{quote} \textit{"Even when virtual reality (VR) was on the rise, our newsroom preferred to outsource that technology. We would temporarily form a cross-functional team to collaborate with VR technologists from tech companies or private studios outside the newsroom to co-produce VR news."} (J2) \end{quote}

AI, however, has significantly changed this dynamic. Unlike data visualization, which journalists can learn through existing programs, AI presents a steep learning curve. Most journalist participants (n = 9) noted that the complexities of AI, such as advanced algorithms and machine learning, exceed their current skills, necessitating deeper collaboration with AI technologists.

\add{This situation also creates a paradox for journalists. Although AI is essential in news distribution and data analysis, most journalists we interviewed find it disconnected from daily work. As J14 noted,} \textit{\add{"I am exploring how generative AI can assist with news writing. It helps organize my thoughts and understand social theories unfamiliar to me, adding depth to my writing."}} \add{J14 also expressed,} \textit{\add{"But do I need to understand machine learning to write news? AI feels distant from my daily work."}} \add{All journalist participants (n=17) echoed that most AI technology feels detached from their routine tasks. J2 shared,} \textit{\add{"Previously, journalists learned data visualization for compelling stories. With AI technologies like NLP, can they genuinely support everyday news writing?"}} \add{This perception discourages AI engagement, worsened by its complexity.}

\add{Nevertheless, machine learning models and algorithms are widely used in journalistic workflows in all newsrooms we interviewed. For example, AI recommends content based on reader interests, uncovers patterns in data analysis, and helps to develop the story by analyzing social media trends. These tasks are always outsourced or managed by in-house AI specialists. According to J13, \textit{"In fact, it is not journalists’ daily news production that needs AI, but the journalism industry that needs AI. AI is needed for audience analysis, big data processing on social networks, and news recommendation system construction. "}}

The deep integration of AI has transformed traditional outsourcing practices into a significant threat to the journalism industry. J3 explained, \textit{"In the past, we would only outsource specific tasks for a particular news project. These collaborations typically lasted for no more than three months and were limited to a single piece of news."} However, AI now permeates almost every aspect of the Chinese news industry, becoming foundational to computational journalism.

J3 elaborated, \textit{"AI is used to extract data from social networks, manage recommendation algorithms, curate news, deliver personalized content, and analyze the audience. Almost every process now involves AI."} This reliance has raised concerns about newsroom autonomy. J13 voiced his apprehension:
\begin{quote} \textit{"Over the past few years, we've collaborated extensively with tech companies. However, they have never adequately explained critical issues such as AI transparency and fairness to us. Even when they tried, almost none of us could understand. Are we supposed to irresponsibly rely on tech companies' AI models in this way?"} (J13)
\end{quote}

This concern was echoed by all journalist participants (n = 17), who agreed that while outsourcing specific projects is acceptable, handing over an entire news process, spanning information gathering, production, distribution, and audience analysis, to tech companies is problematic. They argued that such reliance results in \textit{"the newsroom losing direct control over its content and operations"} (J4), \textit{"a loss of the ability to innovate and adapt to future challenges"} (J6), \textit{"potential misuse, storage, or sharing of data by third-party providers"} (J13), and \textit{"a lack of alignment with the newsroom's unique goals"} (J15).

These two factors, the complexity of AI and outsourcing concerns, have pushed Chinese newsrooms to hire their own AI technologists and build internal cross-functional teams. Given the steep learning curve of AI and the risks of full outsourcing, this shift has become essential for maintaining autonomy and innovation in the industry.

\add{Since journalists are often unwilling or unable to learn complex AI technologies, and outsourcing threatens newsroom autonomy, building internal cross-functional teams and hiring in-house AI technologists is a natural solution. As J5 highlighted,} \textit{\add{"Having AI experts within our organization allows us to maintain editorial control while ensuring the technology aligns with our journalistic goals."}}

\subsection{The Misalignment of AI Design Goals in Collaboration with Technology Companies}

The leading news organizations in China often operate with a dual approach: collaborating with technology companies on the one hand and developing their own solutions on the other. These collaborations with technology companies often involve large-scale projects, such as building NLP-based audience sentiment analysis systems or developing custom GPT models tailored specifically for the news organization. For example, the AI news anchor developed by Xinhua News Agency was created in collaboration with China's leading technology company Sogou ~\cite{kothari2022artificial}. Similarly, CMG Media GPT, developed by China Media Group (CMG), was a joint effort with the Shanghai AI Laboratory ~\cite{shlab2023}. These are typical examples of outsourced projects that consider AI applications in the Chinese news industry.

J5 explained that such large-scale projects, which require substantial human and financial resources, were generally initiated by the Chinese government, as individual newsrooms lacked the capacity to undertake them independently. Typically, the government provides funding, while local technology companies or university laboratories execute the projects in collaboration with newsrooms. According to J5, this has become a common model for developing major AI initiatives in the Chinese news industry. For example,\textit{ "at the provincial level, local governments often partner with regional newspapers and issue tenders to the public or academic institutions to build these projects"} (J5).

However, this outsourcing model often struggles to meet the specific needs of newsrooms. One primary challenge is that newsrooms are increasingly interested in developing AIGC tailored to their daily news production needs. Such content often requires contextualized adjustments. J14 has handled many such projects and provided us with a detailed explanation of the term "contextualized." She pointed out that, \textit{"As journalists, we often start with a vague idea—something that we wish to explore gradually and develop over time. However, government-funded projects typically demand a clear and well-defined objective from the outset, leaving little room for the kind of flexibility."}

Secondly, beyond large-scale, costly AI systems, the newsrooms we looked at were eager to explore a wider range of AI tools that can be applied to various other journalistic tasks, such as news production and information searching. However, these areas often attracted little interest from technology companies because they were not seen as lucrative or high-profile ventures. The lack of commercial appeal and the relatively modest scale of such projects mean that tech companies were less inclined to invest resources in developing tools specifically designed to meet the evolving needs of newsrooms. 

Third, there was a significant difference between how technology companies and newsrooms assess the effectiveness of AI tools. Technology companies often evaluate their products based on metrics favored in computer science, such as algorithmic efficiency. In contrast, newsrooms were more inclined to judge these AI systems or tools based directly on their outputs. This discrepancy resulted in many outsourced products from technology companies failing to meet the specific needs of newsrooms.

According to J5, a lack of communication between technology companies and newsrooms during the outsourcing process exacerbates this issue. According to them, meetings between tech companies and newsrooms were infrequent, and when product reports are presented, they are often difficult for journalists to fully understand. Consequently, journalists in our sample often struggled to provide meaningful feedback for further iterations. While technology companies may demonstrate the effectiveness of their designs through objective data, journalists often remain unsatisfied with the actual use of these tools. For example, they might find that AI-generated text feels too mechanical or that the sentiment analysis results do not contribute to future news production or editorial reviews.

Therefore, according to our participants, cross-functional collaboration between newsrooms and technology companies is often ineffective, with both parties struggling to understand and achieve most of the newsroom's goals. Newsrooms often feel they lack autonomy and meaningful participation in product design when collaborating with technology companies, as these companies are unable to meet all of their design requirements, limiting cooperation to specific products. More importantly, newsrooms sense that their independence and journalistic principles are being eroded by these collaborations. This has led some newsrooms to shift toward internal cross-functional collaboration by hiring their own AI technologists to better align technological development with their specific needs.

\section{~\add{Interview Findings: Current Practices of Cross-Functional Collaboration in the News Industry (RQ1-b)}}\label{practices}

\add{In this section, we explored the key stages of cross-functional collaboration around AI in the Chinese news industry. These collaborations typically encompass five main stages: identifying the collaboration context, defining the task, designing the tool or solve the tasks, applying it in the newsroom, reflecting on cross-functional collaboration.}

\subsection{~\add{Identifying the Collaboration Context}}
\add{Before any specific tasks are defined, cross-functional collaboration requires an overarching understanding of the newsroom's goals and the potential applications of AI. Journalists, particularly those in senior management roles, often outline broad objectives that align with these pressures, exploring opportunities where AI could complement journalistic workflows. In our interviews, newsrooms indicated that they provide corresponding training for both journalists and AI technologists, including lectures and workshops, to familiarize them with the collaborative environment. These sessions often cover topics such as the fundamental applications and prospects of AI in journalism. In some cases, organizations present past collaboration examples to highlight practical use cases and build a shared vision. J7 shared, \textit{"We organize workshops not just to train our team but to create a shared understanding of how AI can be integrated into our workflows."}}

\subsection{~\add{Defining the Specific Task}}
\add{In this stage, the unique editorial priorities of the news industry come to the forefront, as journalists propose task requirements tied to audience engagement, content accuracy, or production efficiency. Senior journalists often initiate this process by articulating high-level objectives, while AI technologists clarify these goals and translate them into achievable engineering tasks. T1 explained, \textit{"Typically, a senior journalist will directly assign us a task, asking us to design a tool or conduct a computational analysis."} T3 added, \textit{"These managers usually have relatively clear high-level objectives, such as needing audience analysis or designing a tool for news source exploration. However, they are often vague about how the analysis should be conducted or the tool designed."} The iterative discussions required to refine these objectives highlight the tension between the editorial goals of journalists and the technical constraints faced by AI technologists. Without alignment during this stage, the risk of creating tools that do not effectively meet newsroom needs increases significantly.}

\subsection{~\add{Designing the Tool or Solving the Specific Computational Task}}
\add{During the design and task solving stage, the focus shifts to balancing technical innovation with journalistic usability. AI technologists always take the lead, according to our participants, as journalists often limit their involvement to providing feedback or monitoring progress. This division of roles reflects the newsroom's emphasis on maintaining editorial integrity while leveraging external technical expertise. J14 highlighted this dynamic: \textit{"It's impossible for someone without expertise to effectively lead those who are experts in their field; doing so inevitably leads to mistakes. That’s why our newsroom has consciously reduced the level of journalist intervention in AI design, allowing AI technologists to take the lead on technical decisions while we focus on defining the broader goals and needs."} While this approach minimizes errors and streamlines technical development, it can result in a lack of iterative feedback from journalists, leading to tools or completed tasks that may not fully align with their expectations or practical needs.}

\subsection{~\add{Applying the Results in the Newsroom}}
\add{This stage focuses on how journalists integrate AI tools or completed computational tasks nto their workflows, but often revealing a disconnect between the tools' design or task-solution intentions and their practical use in real cases we interviewed. For instance, J8 described using a news text generation tool: \textit{"I mainly learned how to operate it. I quickly found it useful for suggesting creative headlines and adjusting content styles for different platforms. That’s all I needed. The technical specifics didn’t matter."} Journalists valued features that improved efficiency but avoided tasks requiring extensive editorial oversight, such as generating full articles. They prioritize editorial quality and credibility over full AI utilization. This is because in applying results from AI practitioners
, news value is always prioritized by journalists.}

\subsection{~\add{Reflections on Cross-Functional Collaboration}}
\add{This stage highlights the reflective process where journalists and AI technologists assess their collaboration and the tools' impact on newsroom practices. Journalists often viewed AI tools as supplementary enhancements rather than critical innovations, prioritizing the preservation of editorial workflows. As J3 reflected, \textit{"As long as AI tools don’t disrupt our workflow, the newsroom won't suffer. We hope they improve areas like distribution, production, and audience analysis, but even if they don’t, it’s not critical."} This cautious stance underscores the newsroom's focus on editorial autonomy and integrity. For AI technologists, this stage often revealed frustrations with misaligned expectations. J10 captured this sentiment: \textit{"Sometimes the AI technologists do meet the initial requirements set by the journalists. But what they create only superficially aligns with those needs without truly fulfilling them. For example, they might provide an analysis of audience opinions, but so what? Just tell me how many comments are positive, neutral, or negative, does that really help me in my news writing? It might be technically impressive, but it's practically useless to us."} These reflections expose the need for iterative learning, better communication, and deeper mutual understanding to bridge the gap between design intentions and practical newsroom needs.}

\add{In summary, cross-functional collaboration in AI within the Chinese news industry is a multi-stage process requiring alignment between journalists' goals and technologists' capabilities. ddressing communication and expectation gaps is critical, as discussed in the next section.}

\section{~\add{Interview and Workshop Findings: Challenges in Internal Cross-Functional Collaborations (RQ2-a)}}\label{challenges}

In previous sections, we highlighted some challenges in cross-functional collaboration.This section provides a structured overview of five key challenges in cross-functional collaboration: (1) journalists and technologists lacked a shared language to reconcile journalistic values with machine logic; (2) power imbalances, where journalists dominated decision-making while technologists were marginalized in newsrooms; and (3) the exclusion of data work, an invisible but vital foundation of collaboration, from cross-functional communication.

\subsection{\newadd{Journalistic Values vs. Machine Logic}}
A major issue in our interviews is the misunderstanding between journalistic values from journalists and machine logic from AI technologists. Technologists often use machine learning jargon that journalists find difficult to understand. In contrast, journalistic concepts such as accuracy, fairness, and value of news are difficult for technologists to grasp. \add{These values are rooted in normative journalism theories that prioritize ethical reporting, public interest, and gatekeeping. For example, "accuracy" involves not just factual correctness but also contextual understanding and editorial judgment. As J3 noted,} \textit{\add{"Accuracy is not just about getting the facts right, it’s about interpreting them in a way that resonates with our audience while staying true to our editorial standards."}} \add{Similarly, "news value" involves subjective assessments of relevance, impact, and timeliness, which vary across newsroom practices and are hard for technologists to operationalize. J15 reflected,} \textit{\add{"We decide what is important not based on algorithms but on human judgment, what our readers care about and what serves the public interest."}}
    
\add{To bridge this gap, AI technologists often reinterpret journalistic concepts into machine-compatible terms. For example, T2 viewed "news distribution" as a recommendation system, focusing on optimizing algorithms to deliver diverse content. This translation helps contextualize journalists' requests within an engineering framework. Similarly, T5 noted,} \textit{\add{"When journalists talk about inclusivity, I think about balancing data inputs to avoid bias in the AI model. It’s a different angle, but it gets us closer to something actionable."}}
    
\add{On the other hand, journalists tend to simplify or bypass technical details, focusing on the content and outcomes of AI tools. J3 remarked,} \textit{\add{"I don’t care how the algorithm works; I just need to know if it can help me write better headlines or summarize content faster. That’s what matters to me."}} \add{According to J3,} \textit{\add{"Editing AI-generated text feels just like editing something written by a junior reporter. You have to clean up the structure, fix inaccuracies, and ensure that it is aligned with the tone and style of the publication. It’s fundamentally no different from editing content written by human journalists in my opinion."}} \add{This pragmatic approach reflects journalists’ focus on editorial utility rather than technical mechanisms. By treating AI tools as workflow extensions, journalists adapt to the technology while minimizing the need for deep technical understanding.}
    
\add{Despite these efforts, the lack of structured frameworks or formal communication processes limits effectiveness, especially in journalism. Unlike industries with defined design processes, journalism relies on interpretative and value-laden decisions, making it hard to align AI capabilities with editorial goals. J7 emphasized,} \textit{\add{"Every piece of news we publish has to reflect a balance between urgency and accuracy; it's not something an algorithm can decide for us."}} \add{These priorities make aligning AI capabilities with editorial goals particularly challenging. Without a shared language, many AI technologists found that even after repeatedly confirming engineering goals with journalists, the final AI tools often disappointed journalists, who felt their needs were not fully understood.}

\add{For instance, T3 shared a case where a journalist requested a tool to "enhance news fairness" but couldn't define "fairness" in measurable terms. T3 remarked,} \textit{\add{"I kept asking, 'What do you mean by fairness? Is it about equal representation, or avoiding bias in the language?' But their responses were always vague, like 'Just make sure it's fair.' It left us guessing how to design the tool."}} \add{Meanwhile, T5 described creating a tool to "reduce selective exposure" as a recommendation algorithm to diversify content consumption. However, according to T5's recollection, the journalist T5 co-worked with later expressed dissatisfaction, saying,} \textit{\add{"We wanted something that encourages readers to engage with diverse perspectives, not just more random recommendations."}} \add{This disconnect highlights the challenge of bridging abstract journalistic ideals with technical execution.}

\newadd{This misalignment is exacerbated by journalists' deep-seated skepticism toward AI, stemming from concerns about quality, accuracy, and political risk. J4 noted,} \textit{\newadd{"AI-generated content often contains numerous errors, so we must meticulously check everything."}} \newadd{Even minor inaccuracies can cause misinformation, risking reputational damage, legal issues, or political scrutiny. } \newadd{In China's regulated media environment, journalists are highly aware of the risks of publishing errors or sensitive content.} \newadd{J6 reflected,} \textit{\newadd{"Our responsibility is not just to our readers but also to the Party and the state. Every piece must align with our values, so we cannot trust AI tools without rigorous oversight."}} 

\newadd{Mistrust is reinforced by the high stakes of content accuracy and ideological alignment in China. Journalists must innovate to meet government expectations yet cannot afford mistakes.} \newadd{J13 stated,} \textit{\newadd{"We want to innovate, but we cannot afford mistakes. The political environment demands perfection, and AI tools aren’t there yet."}} \newadd{This tension, requiring both innovation and extreme caution, fuels industry-wide hesitation toward AI integration.}

\newadd{This dilemma also confuses technologists. T2 admitted struggling to determine whether China’s news industry prioritizes innovation or cautious preservation. For journalists, the path is clearer. J2 explained that news organizations must adopt AI to meet government expectations, often showcasing results during inspections. Yet, even small errors carry catastrophic consequences. J2 likened her newsroom to} \textit{\newadd{"a ship navigating through a storm. We must keep moving forward because standing still means sinking. But if we accelerate recklessly, we risk being destroyed by the storm."}} \newadd{This precarious balance forces both journalists and technologists to navigate the fine line between innovation and caution.}

~\newadd{For the six AI technologists in our study, it was disheartening to learn through our workshops that no journalists had disclosed why the AI tools were ineffective for them previously. They noted that relying on computer science metrics like accuracy and speed did not capture journalists' nuanced expectations for tools that integrate seamlessly into workflows and enhance decision-making. They had suspected this issue but explained, \textit{"Journalists never explained what quality of output they expected from the AI tools; they only mentioned the need for a tool in a certain area, without specifying the desired outcome"} (T3). \textit{"I was left guessing what the journalists wanted and used computer science criteria to create the tools. If the AI product I developed didn't meet their journalistic standards, I don't think that's my fault, but rather a problem with cross-functional communication"} (T5).}

\subsection{Power Imbalance: Dominant Journalists and Marginalized Technologists in Newsrooms}

\add{Our interviews revealed a power imbalance where journalists dominate decision-making, leaving technologists with limited influence. This stems from traditional hierarchies that prioritize editorial authority over technical contributions.} \add{J1 noted, \textit{"Editorial decisions always come first. Technologists support us, not lead or challenge our processes."}} Technologists feel their expertise is undervalued. \add{T4 expressed frustration: \textit{"We're asked to build tools but rarely involved in early discussions. By the time we're included, key decisions are made, leaving little room to suggest alternatives or highlight limitations."}}

\add{This imbalance is further linked to differences in tenure and expertise between journalists and technologists. Experienced journalists often possess extensive institutional knowledge, which grants them authority and trust within the organization. Technologists, on the other hand, are typically recent entrants to the newsroom and may lack the industry-specific expertise needed to assert their perspectives effectively.} \add{T3 noted, \textit{"We understand technology but are still learning journalism. This may contribute to us being seen as supporting players."}} \add{This perception reinforces the notion that technologists are secondary contributors, tasked with execution rather than strategic collaboration. Journalists, meanwhile, may be reluctant to delegate authority to individuals who they feel lack a deep understanding of the complexities and values underpinning journalistic work.} \add{J7 highlighted, \textit{"It's hard to rely on someone who doesn't fully understand our work's nuances. Technologists need time to grasp journalism before taking a central role."}} 

\add{The unequal power discourages open communication. Technologists may hesitate to challenge assumptions, fearing dismissal.} \add{T5 noted, \textit{"There's an unspoken rule that journalists know best. Even if I see a better solution, I hesitate to bring it up."}} \add{This hinders the dialogue necessary for innovative solutions.}

\add{Collaboration thus becomes transactional—focused on tasks rather than integrating expertise from both sides. Technologists are relegated to a supportive role, executing directives without equal involvement in goal-setting or decision-making. This limits meaningful collaboration and reduces opportunities for innovation and workflow improvements.}

\subsection{\add{Invisible Foundations of Cross-Functional Collaboration: Data Work Excluded from Communication}}\label{invisible_labour}

A final challenge identified is the oversight of AI workers in cross-functional collaboration within Chinese newsrooms. \add{In our study, we found these data annotators are crucial for creating and managing labeled datasets essential for AI model development, underpinning the technical processes that enable effective AI systems in newsrooms. However, despite their vital role, they are systematically excluded from communication and decision-making processes.}

\add{Our study recognizes AI workers as integral to cross-functional collaboration, as their efforts sustain the cooperative ecosystem across roles. Although their perspectives are not integrated into communication, including them is essential to achieving AI justice in journalism.} While journalists and AI technologists discuss the design, implementation, and refinement of AI tools, data annotators remain excluded. The three data annotators we interviewed reported never participating in specialized training sessions, formal meetings on AI tool design, or consultations on dataset issues and improvements. This exclusion creates a significant disconnect: their indispensable labor supports AI system development, yet they remain invisible in the collaborative structures shaping AI integration into journalistic workflows.

\add{Nonetheless, T3 and T5 highlighted that by meticulously labeling and annotating vast data, these annotators \textit{"enable AI technologists to train and refine models that automate key journalistic processes"}. Their work ensures the accuracy and effectiveness of AI tools that journalists rely on.}

In our interviews, data annotators like W1 and W2 expressed frustration and confusion. W1 exclaimed, "\textit{I have no idea what data I'm labeling. These are complete BULLSHIT JOBS. Spending 10 hours a day labeling news data makes me furious!}" W2 reflected, "\textit{What is the ultimate purpose of all this data annotation work? No one has ever explained it to us. I don't mind the hard work, but I want to see some practical use and outcomes of my work.}" These AI workers are excluded from career advancement paths available to journalists and technologists. W1, hired at low wages, hoped for an entry point into journalism but found no opportunities to learn or connect with journalists, feeling like outsiders confined to remote work with minimal interaction.

Three AI workers expressed that while they did not expect substantial financial compensation, they valued inclusion in the collaboration process. This desire grew after observing journalists and technologists co-create innovative tools during workshops. W3 explained, \textit{"I took on data annotation to understand newsroom operations and gain skills for my career. Being part of cross-functional teams and learning from journalists and AI experts mattered more to me than the salary."}

\add{Cross-functional collaborations in AI-driven newsrooms also highlight gender disparities tied to power imbalances and invisible labor. AI workers, often engaged in repetitive tasks like data labeling, in our sample, are predominantly female, while AI technologists in strategic roles are mostly male we found. This contrasts sharply with journalism, described as relatively gender-balanced. J17 and W3 noted in WS3, \textit{"AI workers are seen as replaceable, reflecting patterns in traditionally female-dominated roles, while male technologists are viewed as indispensable experts."} J15 added, \textit{"Journalism has made strides in advancing gender equity, but AI-driven roles in newsrooms are bringing back outdated gendered divisions of labor."}}

\add{In summary, data annotators play an invisible yet foundational role in cross-functional collaboration within Chinese newsrooms. Their exclusion from communication and decision-making processes not only undermines their contributions but also hampers the effectiveness and ethical integration of AI tools in journalism. Addressing this exclusion is crucial for fostering inclusive collaboration and achieving AI justice in the news industry.}

\section{\add{Workshop Findings: Future Opportunities in Cross-Functional Collaboration (RQ2-b)}}

\add{This section presents the findings from our workshops, which explored potential opportunities for improving cross-functional collaboration in newsrooms. By bringing together journalists, AI technologists, and AI workers, the workshops provided a platform for participants to collectively reflect on existing challenges, critically examine current practices, and envision future solutions.}

\begin{table*}[h!]
\centering
\caption{\add{Stakeholder Perspectives: Goals, Challenges, and Potential Solutions}}
\label{tab:StakeholderDifferences}
\small
\begin{tabularx}{\textwidth}{lXXX}
\toprule
\textbf{Stakeholder} & \textbf{Goals} & \textbf{Challenges} & \textbf{Potential Solutions} \\
\midrule
Chinese Government & 
Ensuring accurate and compliant information dissemination. Maintaining social stability and a credible but innovative media environment. &
Balancing innovation with regulatory control. Ensuring trust and transparency in AI-driven processes. &
Establishing clear guidelines and standards. Emphasizing Incremental Innovation. Utilizing Pilot Projects. Balancing Innovation and Political Sensitivities.\\[1em]

Journalists &
Producing high-quality, timely, and trustworthy stories. Leveraging AI tools while retaining editorial independence. &
Limited understanding of AI capabilities. Concerns over automation replacing human judgment. Difficulty integrating AI insights into daily workflow. &
Providing targeted training on AI tools. Involving journalists in early tool design stages. \\[1em]

AI Technologists &
Developing reliable, user-friendly AI tools for the newsroom. Efficiently incorporating editorial feedback. Ensuring data quality and relevance. &
Gaps in domain knowledge regarding journalism practices. Difficulty securing quality annotated data. Adapting tools to evolving newsroom needs. &
Conducting collaborative sessions to align technical and editorial goals. Defining data requirements with journalists and workers. Engaging in iterative prototype development. Piloting and refining tools in real reporting contexts.\\[1em]

AI Workers &
Undertaking clearly defined annotation tasks. Improving skills and advancing professionally. Enjoying fair working conditions. &
Repetitive and sometimes unclear labeling tasks. Limited career growth and lack of recognition. &
Offering skill development programs and internal workshops. Recognizing contributions and ensuring stable contracts.\\
\bottomrule
\end{tabularx}
\end{table*}

\subsection{\add{Transform Abstract Concepts into Concrete Ones: Translating Journalistic Theory into Machine Terms}}

\add{Workshops revealed that effective cross-functional collaboration requires clear and specific communication, avoiding vague terms. When concepts are unclear, participants often use analogies to bridge understanding. For instance, several (n=4) participants used analogies to translate each other’s concepts. J13 explained in WS1,} \textit{\add{"When they talk about training data or algorithms, I try to imagine it as an editorial process, it’s the only way I can make sense of it."}} \add{T3 also reflected,} \textit{\add{"When journalists describe 'news fairness,' I think of it as balancing weights in a model, it’s how I translate their ideas into something actionable."}}

\add{Collaboration goes beyond teaching AI to journalists or vice versa. J4 and T4 observed,} \textit{\add{"It’s not just about knowing how AI works—it’s about integrating it into our workflows, values, and news production."}} \add{J4 added,} \textit{\add{"When we design a tool, we’re testing how it can push the boundaries of journalism."}} \add{AI expands news boundaries, complicating precise descriptions.}

\add{Participants proposed practical strategies to translate abstract concepts into concrete terms. One was creating \textit{cross-disciplinary glossaries} to define key terms for clearer communication. J3 suggested,} \textit{\add{"If we agree on what 'fairness' or 'engagement' means from both perspectives, it would eliminate a lot of the initial confusion."}} \add{Another solution was \textit{collaborative prototyping sessions}, where journalists and technologists co-create tools or workflows, allowing iterative feedback. T3 remarked,} \textit{\add{"Seeing a rough prototype or flowchart helps journalists articulate their needs more clearly than abstract discussions."}}

\add{Participants emphasized structured training sessions to foster mutual understanding, not just technical skills. These could include scenario-based learning, where journalists and technologists collaborate on hypothetical cases to explore AI and journalistic values intersections. Such methods enhance technical literacy and strengthen relational dynamics. J15 noted,} \textit{\add{"The more we understand each other’s work, the easier it becomes to collaborate effectively."}}

\add{Participants envisioned a future newsroom where concept translation is automated by advanced AI interfaces. An\textit{ AI Translation Engine }could convert journalistic values like "fairness" into actionable design parameters for technologists. J13 speculated in WS2 and WS3,} \textit{\add{"What if we could input 'fairness' and the system shows metrics, datasets, and design options? It could save months of trial and error."}} \add{Additionally, a \textit{Collaborative Design Simulator} could allow real-time visualization of how design choices impact workflows and audience reception. T6 remarked,} \textit{\add{"It’s not just about understanding terms, it’s about seeing how our work connects and what the outcomes look like before we start building."}}

\subsection{\add{Enhancing Cross-Functional Collaboration Through Co-Design and Prototyping}}

\add{All participants highlighted persistent communication gaps between journalists and AI technologists in cross-functional collaboration. During workshops, they explored iterative and tangible design methods to bridge these gaps. For instance, a co-design approach encouraged AI technologists to develop simple, low-fidelity prototypes based on journalists’ initial needs. Journalists could then test these prototypes, providing feedback for rapid refinement or abandoning ineffective ideas. This iterative process fosters faster alignment and reduces wasted effort on both sides.}

\add{For journalists who often struggle with abstract AI concepts, interacting with prototype tools made AI capabilities more concrete and accessible. As J15 remarked in the second workshop, \textit{"Seeing the prototype in action makes it much easier to understand what’s possible and what’s not. It helps us figure out what we really need."} This hands-on approach clarifies journalists' requirements, creating a feedback loop that supports ongoing discussions with AI technologists. Similarly, iterative feedback helps technologists adapt their designs to better meet journalistic needs without overinvesting in misaligned ideas.}

\add{Participants also emphasized contextualizing prototypes within the newsroom's specific environment. T4 noted, \textit{"It’s not enough to just build something, we need to show how it fits into their day-to-day work."} In the imagination of our participants, by situating prototypes in real-world scenarios, teams can explore how AI tools align with values like accuracy, fairness, and audience engagement. This approach could address the challenge that journalists often struggle to envision AI tools' purpose or benefits without seeing their functionality in the newsroom.}

\add{All participants (n=9) agreed that future HCI and journalism research should focus on designing methods and tools that integrate prototypes into cross-functional journalistic workflows. They emphasized creating platforms for real-time collaboration. J4 suggested, \textit{"We need a space where we can see the evolution of a tool, how our feedback changes it and how it will fit into our work."} Similarly, T3 proposed, \textit{"What if we had a dashboard that lets us test ideas together in real time? It would save us from miscommunications and wasted effort."}}

\add{According to our workshops, participants argue that future HCI tools for journalism could establish shared workflows, ensuring journalists remain actively involved throughout AI tool development stages. By enabling ongoing interaction and feedback, these platforms could bridge disciplinary divides and foster a culture of shared accountability. As J3 concluded, \textit{"If we can keep working together at every stage, the tools we build will actually reflect what both sides need."}}

\subsection{\add{Inclusive Cross-Functional Collaboration: Balancing Power and Addressing Invisible Labor}}

\add{Cross-functional collaborations in AI-driven newsrooms face two key issues: power imbalances between journalists and AI technologists, and the invisibility of AI workers. These challenges hinder inclusivity and limit innovative collaboration.}

\add{Participants highlighted that in current newsrooms, journalists dominate decision-making, relegating technologists to secondary roles. J13 reflected, \textit{"We used to see technologists as instrumental roles, only helping us meet our needs. This might have worked in data journalism, but it's outdated in the AI era. Since we don't fully understand what LLMs can do, we must incorporate AI experts into a multi-role newsroom framework."}}

\add{Moreover, AI workers face greater exclusion. W3 described, \textit{"We handle the data that powers these systems, but we’re kept at a distance, no one explains why our work matters or how it’s used."} This lack of transparency undermines their recognition and deprives them of learning opportunities. J15 remarked, \textit{"If we bring AI workers into the newsroom, they can better understand their work's purpose and contribute ideas we might not have considered."}}

\add{Participants in WS3 further viewed gender and power imbalances as intertwined, creating intersectional inequalities. Then, they suggested integrating AI workers into collaborative newsroom workflows through mentorship programs and equitable access to professional development opportunities could help elevate their visibility and contributions. T4 proposed, \textit{"We could have mentorship systems that focus on skills transfer and upward mobility for AI workers, especially women, to balance the scales."}}

\add{To address these challenges, participants proposed practical strategies. First, integrating AI workers into newsroom workflows through structured training, cross-role mentorship, and collaborative workshops would enhance their professional development and improve AI tool quality. Second, increasing transparency around data processes could foster shared accountability. T4 suggested, \textit{"Making the data process visible would help everyone see its value and ensure alignment with our shared goals."} Finally, establishing shared workflows can dismantle hierarchical power dynamics. T2 proposed, \textit{"Everyone in the newsroom must respect each other. Anyone in journalism can be a journalist, including AI technologists and workers."}}

\add{Participants envisioned a future newsroom with restructured power relations to foster genuine inclusivity and collaboration. In this model, traditional hierarchies would give way to an equitable system where all roles have an equal voice in decision-making. T4 proposed, \textit{"We need systems that redistribute power, ensuring technologists and data workers are active contributors to journalism."} J17 elaborated, \textit{"Imagine a newsroom where everyone, regardless of role, has equal access to resources and decision-making spaces. It’s about redefining what it means to be part of journalism."}}

\add{Ultimately, in envision of all our workshops participants (N=9), this speculative newsroom represents a radical departure from current norms, prioritizing shared responsibility and mutual respect. By embedding equality into workflows, this vision aims to create a collaborative culture where all voices carry equal weight, unlocking the full potential of cross-functional collaboration in journalism.}

\section{Discussion}

\add{At the CHI '19 workshop on "HCI for Accurate, Impartial, and Transparent Journalism," HCI scholars outlined key principles for HCI's role in journalism~\cite{aitamurto2019hci}, emphasizing the importance of designing technologies that align with journalistic values, support transparency, and enhance trust.} \add{Our study advances the discourse on HCI in journalism by addressing the integration challenges of AI in newsrooms during cross-functional collaboration, particularly within China's socio-political landscape. Building on the CHI '19 principles~\cite{aitamurto2019hci}, we investigate how cross-functional collaboration among journalists, technologists, and AI workers can align with journalistic values while managing governmental pressures. By situating our findings within design theory, HCI frameworks, and journalism theory, we offer an interdisciplinary perspective that bridges gaps between these roles, fostering innovations consistent with journalistic integrity.} \add{We now outline our contributions to HCI theory, future HCI design, and journalism research.}

\subsection{\add{Bridging News Values and Cross-Functional Collaboration Through HCI Principles}}

\add{While existing studies emphasize shared goals and mutual understanding in cross-functional collaboration~\cite{jassawalla1999building,dodds2024impact,majchrzak2012transcending}, our research identifies unique barriers in journalism, such as power imbalances, invisible roles (e.g., AI workers), and the absence of a shared language aligning technical and journalistic values. These insights enrich the discourse on cross-functional teamwork in value-laden~\cite{cronin2019conflict} and politically sensitive environments~\cite{simmons2022political}.}

\add{A key contribution is the critical examination of power asymmetries. Traditional theories assume equitable collaboration~\cite{salvato2017cooperation,deng2023investigating,deng2023understanding}, but our study reveals how newsroom hierarchies marginalize technologists and AI workers. By detailing mechanisms like exclusion from early discussions and limited recognition, we provide a deep understanding of how power imbalances disrupt collaboration, expanding theories to include structural barriers~\cite{bernstein2020diversity}.}

\add{Additionally, we highlight the necessity of bridging conceptual gaps~\cite{majchrzak2012transcending,deng2023investigating,agapie2024conducting}. Cross-functional collaboration theories often overlook how abstract journalistic values, such as fairness or efficiency, complicate interdisciplinary alignment. Our findings demonstrate that tools like prototypes and boundary objects can translate these abstract concepts into actionable goals. Workshops showed that iterative prototyping not only facilitates communication but also clarifies mutual expectations, enhancing theoretical frameworks by emphasizing tangible artifacts in interdisciplinary teams~\cite{nicolini2012understanding}.}

\add{Furthermore, we emphasize iterative feedback mechanisms as essential for effective collaboration~\cite{d2005conceptual,popoola2024conceptualizing,yang2020re}. Unlike linear models, our findings suggest that iterative loops, where outputs are continuously evaluated and refined, are crucial for dynamic teamwork. This approach ensures tools evolve to meet stakeholders' needs, fostering mutual trust and accountability~\cite{ayobi2023computational}.}

\add{In summary, our contributions include (1) critiquing power imbalances that undermine equitable teamwork, (2) proposing mechanisms to translate abstract values into concrete goals, and (3) emphasizing iterative feedback as a core principle for sustaining collaboration. These insights enhance our understanding of cross-functional collaboration and offer practical pathways for more inclusive and dynamic collaborative practices in journalism.}

\subsection{\add{Designing for Journalistic Inclusivity: Prototyping Tools for Cross-Functional Collaboration}}

\add{Our research underscores the importance of prototyping tools that facilitate collaboration in journalism and suggests future HCI tools to support journalistic work, thus contributing to HCI design and computer-supported cooperative work studies. Traditional newsroom workflows prioritize editorial authority~\cite{cook2012governing}, but co-designed tools can bridge communication gaps between journalists, technologists, and AI workers. We offer actionable suggestions for future HCI design in newsrooms.}

\add{Key contributions from our workshops include the development of mediating artifacts like interactive dashboards and rapid prototyping platforms~\cite{yao2024exploring}. These tools enable stakeholders to collaboratively test and refine AI-driven solutions, aligning diverse expertise early in the design process~\cite{dove2017ux,yang2020re}. For example, participants envisioned prototypes that translate journalistic values into actionable design parameters, making abstract concepts accessible to technologists.}

\add{Additionally, iterative feedback is crucial for refining AI tools for journalism. Journalists found that interacting with prototypes helped clarify their requirements, creating a feedback loop that supports ongoing dialogue with technologists. This aligns with previous studies suggesting that prototypes help journalists identify missing or misaligned elements~\cite{wagemans2019examining,aitamurto2023future}.}

\add{Our findings also suggest that prototyping tools can address power imbalances by democratizing the design process~\cite{costanza2020design}. By making contributions from all roles visible, these tools ensure equitable recognition and participation. This aligns with HCI’s goal of designing inclusive, value-driven systems~\cite{sadek2023designing}. Future research could explore scaling these tools to support larger news organizations, fostering sustainable cross-disciplinary collaboration.}

\add{Ultimately, our work advances HCI’s focus on user-centered, participatory systems. By integrating prototyping tools into newsrooms, we demonstrate how HCI methodologies can empower stakeholders to co-create solutions that reflect both journalistic values and technical feasibility, driving innovation while maintaining inclusivity~\cite{aitamurto2019hci}.}

\subsection{\add{Trading Zones and Cross-Functional Collaboration about AI in Journalism}}

\add{Cross-functional collaboration in journalism operates within "trading zones"~\cite{galison2010trading,shrager2010wizards}, shared spaces where individuals negotiate meanings and translate concepts to achieve collective goals. These zones enable effective communication despite diverse disciplinary perspectives~\cite{lewis2016trading}. For journalism, trading zones integrate journalistic values, fairness, transparency, credibility, into AI tool design, ensuring alignment with editorial priorities and public accountability.}

\add{Our study highlights challenges within these trading zones, such as power imbalances and communication barriers. Journalist-centric decision-making limits technologists’ and AI workers’ contributions, reducing trading zones to technical support rather than collaborative co-creation~\cite{shrager2010wizards,dodds2024collaborative}. Additionally, the lack of a shared language exacerbates these issues. Journalists frame requirements in editorial terms, while technologists use technical jargon. Although analogies and informal translations help, they lack a systematic structure for mutual understanding, preventing trading zones from fully facilitating interdisciplinary collaboration.}

\add{To overcome these challenges, we propose formalizing trading zones with tools and workflows that foster mutual understanding. Translating journalistic values into measurable design parameters and including all stakeholders from the outset can transform trading zones into equitable, integrative spaces. Our findings expand the trading zones framework by addressing structural inequalities and aligning disciplinary vocabularies, bridging divides that hinder innovation and fostering inclusivity in cross-functional collaborations.}

\subsection{\add{Reimagining Future Journalism: Integrating AI and Redefining Newsroom Dynamics}}

\add{Our research contributes to journalism studies by challenging traditional newsroom hierarchies and advocating for an inclusive collaboration model~\cite{walters2024new}. Traditionally, the editorial authority has been central to news production, sidelining technologists and data workers~\cite{cook2012governing}. However, AI's growing role necessitates a shift toward more equitable newsroom dynamics.} \add{We advocate for restructuring power dynamics to include technologists and AI workers in strategic discussions and design processes. This integration allows technologists to contribute their expertise, creating tools better aligned with editorial goals. }

\add{Our findings emphasize the interdependence between journalistic values and AI capabilities~\cite{gutierrez2023question}. Journalists uphold ethical standards and content quality, while technologists provide AI system capabilities. Aligning these perspectives allows the co-creation of tools that enhance credibility and audience engagement without compromising editorial integrity. For example, iterative prototyping demonstrated how mutual understanding can translate abstract principles into actionable technical goals.}

\add{This study aligns with Aitamurto et al.'s manifesto~\cite{aitamurto2023future}, which calls for integrating design methodologies into journalism to navigate digital transformation. We extend this vision by illustrating how participatory and speculative design could enhance collaboration in AI-driven newsrooms while stimulating collective exploration of AI justice in newsrooms.}

\add{Finally, our study underscores the importance of balancing innovation with caution in politically sensitive environments like China~\cite{nguyen2022new}. Newsrooms in China must adopt AI to meet governmental expectations while maintaining strict standards of accuracy and credibility to preserve public trust and avoid political repercussions. Embedding iterative feedback and transparent evaluation criteria into AI workflows mitigates these tensions, ensuring technological advancements support journalistic integrity.}

\add{Our work offers a theoretical and practical framework for integrating AI into journalism in innovative and inclusive ways that align with editorial priorities. By redefining power dynamics and enhancing cross-functional collaboration, we help newsrooms harness AI's transformative potential while upholding core journalistic values. As AI reshapes journalism, our focus on collaboration between AI specialists and journalists provides actionable insights for seamless integration into workflows. This perspective goes beyond discussing AI's challenges and opportunities, highlighting how collaboration can adapt and preserve journalistic values. This approach advances both journalism theory and practice in the AI era.}

\section{Limitations and Future Work}

Our study has several limitations. First, it focuses solely on China, limiting the generalizability of findings due to its unique socio-political context. Future research should explore AI integration in diverse geopolitical settings. Second, we examined only leading news organizations, overlooking smaller outlets that may face different challenges and opportunities. Third, our focus on internal newsroom collaboration excludes interactions with external tech companies, which play a critical role in AI adoption. Fourth, we studied only data annotators among AI workers, neglecting other roles like content moderators. Expanding future research to include diverse organizations, roles, and external partnerships, as well as integrating organizational behavior theories and mixed-methods approaches, could provide a more comprehensive understanding of cross-functional collaboration in journalism.

\section{Conclusion}

This study examines strategies and challenges in cross-functional collaboration around AI in the Chinese news industry, aiming to identify improvements for the field. Interviews and workshops with Chinese journalists, AI technologists, and AI workers revealed that due to AI's unique characteristics, outsourcing-driven autonomy crises, and political tensions, the industry tends to form effective internal cross-functional teams. We explored how collaboration unfolds across AI tool design stages in journalism and how interdisciplinary knowledge is integrated, highlighting key challenges. This work aims to enhance practitioners' understanding of AI collaboration and contribute empirically and theoretically to HCI research on cross-functional collaboration.

\bibliographystyle{ACM-Reference-Format}
\bibliography{references}


\begin{thebibliography}{109}


\ifx \showCODEN    \undefined \def \showCODEN     #1{\unskip}     \fi
\ifx \showDOI      \undefined \def \showDOI       #1{#1}\fi
\ifx \showISBNx    \undefined \def \showISBNx     #1{\unskip}     \fi
\ifx \showISBNxiii \undefined \def \showISBNxiii  #1{\unskip}     \fi
\ifx \showISSN     \undefined \def \showISSN      #1{\unskip}     \fi
\ifx \showLCCN     \undefined \def \showLCCN      #1{\unskip}     \fi
\ifx \shownote     \undefined \def \shownote      #1{#1}          \fi
\ifx \showarticletitle \undefined \def \showarticletitle #1{#1}   \fi
\ifx \showURL      \undefined \def \showURL       {\relax}        \fi
\providecommand\bibfield[2]{#2}
\providecommand\bibinfo[2]{#2}
\providecommand\natexlab[1]{#1}
\providecommand\showeprint[2][]{arXiv:#2}

\bibitem[Agapie et~al\mbox{.}(2024)]%
        {agapie2024conducting}
\bibfield{author}{\bibinfo{person}{Elena Agapie}, \bibinfo{person}{Ravi Karkar}, \bibinfo{person}{Tricia Aung}, \bibinfo{person}{Eleanor~R Burgess}, \bibinfo{person}{Munyaradzi~Joel Chinguwa}, \bibinfo{person}{Andrea~K Graham}, \bibinfo{person}{Predrag Klasnja}, \bibinfo{person}{Aaron Lyon}, \bibinfo{person}{Terika McCall}, \bibinfo{person}{Sean~A Munson}, {et~al\mbox{.}}} \bibinfo{year}{2024}\natexlab{}.
\newblock \showarticletitle{Conducting Research at the Intersection of HCI and Health: Building and Supporting Teams with Diverse Expertise to Increase Public Health Impact}. In \bibinfo{booktitle}{\emph{Extended Abstracts of the CHI Conference on Human Factors in Computing Systems}}. \bibinfo{pages}{1--6}.
\newblock


\bibitem[Aitamurto et~al\mbox{.}(2019)]%
        {aitamurto2019hci}
\bibfield{author}{\bibinfo{person}{Tanja Aitamurto}, \bibinfo{person}{Mike Ananny}, \bibinfo{person}{Chris~W Anderson}, \bibinfo{person}{Larry Birnbaum}, \bibinfo{person}{Nicholas Diakopoulos}, \bibinfo{person}{Matilda Hanson}, \bibinfo{person}{Jessica Hullman}, {and} \bibinfo{person}{Nick Ritchie}.} \bibinfo{year}{2019}\natexlab{}.
\newblock \showarticletitle{HCI for accurate, impartial and transparent journalism: Challenges and solutions}. In \bibinfo{booktitle}{\emph{Extended abstracts of the 2019 CHI conference on human factors in computing systems}}. \bibinfo{pages}{1--8}.
\newblock


\bibitem[Aitamurto et~al\mbox{.}(2023)]%
        {aitamurto2023future}
\bibfield{author}{\bibinfo{person}{Tanja Aitamurto}, \bibinfo{person}{Eddy Borges-Rey}, {and} \bibinfo{person}{Nicholas Diakopoulos}.} \bibinfo{year}{2023}\natexlab{}.
\newblock \bibinfo{title}{The future of design+ journalism: a manifesto for bridging digital journalism and design}.
\newblock , \bibinfo{numpages}{399--410}~pages.
\newblock


\bibitem[Akpe et~al\mbox{.}(2024)]%
        {akpe2024immersive}
\bibfield{author}{\bibinfo{person}{Samuel Akpe}, \bibinfo{person}{Jan~Lauren Boyles}, \bibinfo{person}{Kaixin Cheng}, \bibinfo{person}{Andrea Hudson}, \bibinfo{person}{Samantha-Kaye Johnston}, \bibinfo{person}{Seok Kang}, \bibinfo{person}{Zixuan Li}, \bibinfo{person}{Tijmen Koppelaar}, \bibinfo{person}{Desmond~Onyemechi Okocha}, \bibinfo{person}{Ivanka Pjesivac}, {et~al\mbox{.}}} \bibinfo{year}{2024}\natexlab{}.
\newblock \bibinfo{booktitle}{\emph{Immersive Journalism: Virtual Reality and the Future of the News Industry}}.
\newblock \bibinfo{publisher}{Rowman \& Littlefield}.
\newblock


\bibitem[Aljalabneh et~al\mbox{.}(2024)]%
        {aljalabneh2024balancing}
\bibfield{author}{\bibinfo{person}{Abdallah Aljalabneh}, \bibinfo{person}{Hamzeh Aljawawdeh}, \bibinfo{person}{Alia Mahmoud}, \bibinfo{person}{Tahseen Sharadqa}, {and} \bibinfo{person}{Ashraf Al-Zoubi}.} \bibinfo{year}{2024}\natexlab{}.
\newblock \showarticletitle{Balancing Efficiency and Ethics: The Challenges of Artificial Intelligence Implementation in Journalism}.
\newblock \bibinfo{journal}{\emph{Intelligent Systems, Business, and Innovation Research}} (\bibinfo{year}{2024}), \bibinfo{pages}{763--773}.
\newblock


\bibitem[Almahmoud et~al\mbox{.}(2021)]%
        {almahmoud2021teams}
\bibfield{author}{\bibinfo{person}{Jumana Almahmoud}, \bibinfo{person}{Robert DeLine}, {and} \bibinfo{person}{Steven~M Drucker}.} \bibinfo{year}{2021}\natexlab{}.
\newblock \showarticletitle{How teams communicate about the quality of ML models: a case study at an international technology company}.
\newblock \bibinfo{journal}{\emph{Proceedings of the ACM on Human-Computer Interaction}} \bibinfo{volume}{5}, \bibinfo{number}{GROUP} (\bibinfo{year}{2021}), \bibinfo{pages}{1--24}.
\newblock


\bibitem[Ambos et~al\mbox{.}(2013)]%
        {ambos2013combining}
\bibfield{author}{\bibinfo{person}{Tina~C Ambos}, \bibinfo{person}{Phillip~C Nell}, {and} \bibinfo{person}{Torben Pedersen}.} \bibinfo{year}{2013}\natexlab{}.
\newblock \showarticletitle{Combining stocks and flows of knowledge: The effects of intra-functional and cross-functional complementarity}.
\newblock \bibinfo{journal}{\emph{Global Strategy Journal}} \bibinfo{volume}{3}, \bibinfo{number}{4} (\bibinfo{year}{2013}), \bibinfo{pages}{283--299}.
\newblock


\bibitem[Aronson-Rath et~al\mbox{.}(2015)]%
        {aronson2015virtual}
\bibfield{author}{\bibinfo{person}{R. Aronson-Rath}, \bibinfo{person}{J. Milward}, \bibinfo{person}{T. Owen}, {and} \bibinfo{person}{F. Pitt}.} \bibinfo{year}{2015}\natexlab{}.
\newblock \bibinfo{booktitle}{\emph{Virtual Reality Journalism}}.
\newblock \bibinfo{type}{Research Project}. \bibinfo{institution}{The Tow Center for Digital Journalism at Columbia University}.
\newblock
\newblock
\shownote{A Research Project by The Tow Center for Digital Journalism at Columbia University}.


\bibitem[Ayobi et~al\mbox{.}(2023)]%
        {ayobi2023computational}
\bibfield{author}{\bibinfo{person}{Amid Ayobi}, \bibinfo{person}{Jacob Hughes}, \bibinfo{person}{Christopher~J Duckworth}, \bibinfo{person}{Jakub~J Dylag}, \bibinfo{person}{Sam James}, \bibinfo{person}{Paul Marshall}, \bibinfo{person}{Matthew Guy}, \bibinfo{person}{Anitha Kumaran}, \bibinfo{person}{Adriane Chapman}, \bibinfo{person}{Michael Boniface}, {et~al\mbox{.}}} \bibinfo{year}{2023}\natexlab{}.
\newblock \showarticletitle{Computational notebooks as co-design tools: engaging young adults living with diabetes, family carers, and clinicians with machine learning models}. In \bibinfo{booktitle}{\emph{Proceedings of the 2023 CHI Conference on Human Factors in Computing Systems}}. \bibinfo{pages}{1--20}.
\newblock


\bibitem[{BBC News}(2024)]%
        {bbc_ai_guidance_2024}
\bibfield{author}{\bibinfo{person}{{BBC News}}.} \bibinfo{year}{2024}\natexlab{}.
\newblock \bibinfo{title}{Guidance: The use of Artificial Intelligence}.
\newblock
\newblock
\urldef\tempurl%
\url{https://www.bbc.co.uk/editorialguidelines/guidance/use-of-artificial-intelligence}
\showURL{%
\tempurl}
\newblock
\shownote{Accessed: 2024-07-22}.


\bibitem[Becker et~al\mbox{.}(2023)]%
        {simon2023policies}
\bibfield{author}{\bibinfo{person}{Kim~Björn Becker}, \bibinfo{person}{Felix~M Simon}, {and} \bibinfo{person}{Christopher Crum}.} \bibinfo{year}{2023}\natexlab{}.
\newblock \showarticletitle{Policies in parallel? A comparative study of journalistic AI policies in 52 global news organisations}.
\newblock  (\bibinfo{year}{2023}).
\newblock


\bibitem[Bernstein et~al\mbox{.}(2020)]%
        {bernstein2020diversity}
\bibfield{author}{\bibinfo{person}{Ruth~Sessler Bernstein}, \bibinfo{person}{Morgan Bulger}, \bibinfo{person}{Paul Salipante}, {and} \bibinfo{person}{Judith~Y Weisinger}.} \bibinfo{year}{2020}\natexlab{}.
\newblock \showarticletitle{From diversity to inclusion to equity: A theory of generative interactions}.
\newblock \bibinfo{journal}{\emph{Journal of Business Ethics}}  \bibinfo{volume}{167} (\bibinfo{year}{2020}), \bibinfo{pages}{395--410}.
\newblock


\bibitem[{Bloomberg}(2023)]%
        {Bloomberg2023}
\bibfield{author}{\bibinfo{person}{{Bloomberg}}.} \bibinfo{year}{2023}\natexlab{}.
\newblock \bibinfo{title}{Introducing BloombergGPT, Bloomberg’s 50-billion parameter large language model, purpose-built from scratch for finance}.
\newblock
\newblock
\urldef\tempurl%
\url{https://www.bloomberg.com/company/press/bloomberggpt-50-billion-parameter-llm-tuned-finance/}
\showURL{%
\tempurl}
\newblock
\shownote{Accessed: 2024-09-10}.


\bibitem[Boivin and CohenMiller(2018)]%
        {boivin2018breaking}
\bibfield{author}{\bibinfo{person}{Nettie Boivin} {and} \bibinfo{person}{Anna CohenMiller}.} \bibinfo{year}{2018}\natexlab{}.
\newblock \showarticletitle{Breaking the “fourth wall” in qualitative research: Participant-led digital data construction}.
\newblock \bibinfo{journal}{\emph{Qualitative Report}} \bibinfo{volume}{23}, \bibinfo{number}{3} (\bibinfo{year}{2018}).
\newblock


\bibitem[Broussard et~al\mbox{.}(2019)]%
        {broussard2019artificial}
\bibfield{author}{\bibinfo{person}{Meredith Broussard}, \bibinfo{person}{Nicholas Diakopoulos}, \bibinfo{person}{Andrea~L Guzman}, \bibinfo{person}{Rediet Abebe}, \bibinfo{person}{Michel Dupagne}, {and} \bibinfo{person}{Ching-Hua Chuan}.} \bibinfo{year}{2019}\natexlab{}.
\newblock \showarticletitle{Artificial intelligence and journalism}.
\newblock \bibinfo{journal}{\emph{Journalism \& mass communication quarterly}} \bibinfo{volume}{96}, \bibinfo{number}{3} (\bibinfo{year}{2019}), \bibinfo{pages}{673--695}.
\newblock


\bibitem[Cai et~al\mbox{.}(2021)]%
        {cai2021onboarding}
\bibfield{author}{\bibinfo{person}{Carrie~J Cai}, \bibinfo{person}{Samantha Winter}, \bibinfo{person}{David Steiner}, \bibinfo{person}{Lauren Wilcox}, {and} \bibinfo{person}{Michael Terry}.} \bibinfo{year}{2021}\natexlab{}.
\newblock \showarticletitle{Onboarding Materials as Cross-functional Boundary Objects for Developing AI Assistants}. In \bibinfo{booktitle}{\emph{Extended Abstracts of the 2021 CHI Conference on Human Factors in Computing Systems}}. \bibinfo{pages}{1--7}.
\newblock


\bibitem[Cai and Nishal(2023)]%
        {cai2023motivations}
\bibfield{author}{\bibinfo{person}{Mandi Cai} {and} \bibinfo{person}{Sachita Nishal}.} \bibinfo{year}{2023}\natexlab{}.
\newblock \showarticletitle{Motivations, Goals, and Pathways for AI Literacy for Journalism}. In \bibinfo{booktitle}{\emph{CHI’23 Workshop on AI Literacy: Finding Common Threads between Education, Design, Policy, and Explainability}}.
\newblock


\bibitem[Caswell(2023)]%
        {caswell2023ai}
\bibfield{author}{\bibinfo{person}{David Caswell}.} \bibinfo{year}{2023}\natexlab{}.
\newblock \showarticletitle{AI and journalism: What’s next}.
\newblock \bibinfo{journal}{\emph{Reuters Institute}} (\bibinfo{year}{2023}).
\newblock


\bibitem[{Central South University}(2024)]%
        {csu2024generativeAI}
\bibfield{author}{\bibinfo{person}{{Central South University}}.} \bibinfo{year}{2024}\natexlab{}.
\newblock \bibinfo{title}{Generative Artificial Intelligence and Journalism Communication Graduate Summer School Held at Central South University}.
\newblock
\newblock
\urldef\tempurl%
\url{https://clj.csu.edu.cn/info/1054/6693.htm}
\showURL{%
\tempurl}
\newblock
\shownote{Accessed: 2024-09-08}.


\bibitem[Chase(2017)]%
        {chase2017}
\bibfield{author}{\bibinfo{person}{Elizabeth Chase}.} \bibinfo{year}{2017}\natexlab{}.
\newblock \showarticletitle{Enhanced Member Checks: Reflections and Insights from a Participant-Researcher Collaboration}.
\newblock \bibinfo{journal}{\emph{The Qualitative Report}} \bibinfo{volume}{22}, \bibinfo{number}{10} (\bibinfo{year}{2017}), \bibinfo{pages}{2718--2730}.
\newblock
\urldef\tempurl%
\url{https://core.ac.uk/download/pdf/132324724.pdf}
\showURL{%
\tempurl}


\bibitem[Chinh et~al\mbox{.}(2019)]%
        {chinh2019ways}
\bibfield{author}{\bibinfo{person}{Bonnie Chinh}, \bibinfo{person}{Himanshu Zade}, \bibinfo{person}{Abbas Ganji}, {and} \bibinfo{person}{Cecilia Aragon}.} \bibinfo{year}{2019}\natexlab{}.
\newblock \showarticletitle{Ways of qualitative coding: A case study of four strategies for resolving disagreements}. In \bibinfo{booktitle}{\emph{Extended abstracts of the 2019 CHI conference on human factors in computing systems}}. \bibinfo{pages}{1--6}.
\newblock


\bibitem[Collins et~al\mbox{.}(2007)]%
        {collins2007trading}
\bibfield{author}{\bibinfo{person}{Harry Collins}, \bibinfo{person}{Robert Evans}, {and} \bibinfo{person}{Mike Gorman}.} \bibinfo{year}{2007}\natexlab{}.
\newblock \showarticletitle{Trading zones and interactional expertise}.
\newblock \bibinfo{journal}{\emph{Studies in History and Philosophy of Science Part A}} \bibinfo{volume}{38}, \bibinfo{number}{4} (\bibinfo{year}{2007}), \bibinfo{pages}{657--666}.
\newblock


\bibitem[Cook(2012)]%
        {cook2012governing}
\bibfield{author}{\bibinfo{person}{Timothy~E Cook}.} \bibinfo{year}{2012}\natexlab{}.
\newblock \bibinfo{booktitle}{\emph{Governing with the news: The news media as a political institution}}.
\newblock \bibinfo{publisher}{University of Chicago press}.
\newblock


\bibitem[Costanza-Chock(2020)]%
        {costanza2020design}
\bibfield{author}{\bibinfo{person}{Sasha Costanza-Chock}.} \bibinfo{year}{2020}\natexlab{}.
\newblock \bibinfo{booktitle}{\emph{Design justice: Community-led practices to build the worlds we need}}.
\newblock \bibinfo{publisher}{The MIT Press}.
\newblock


\bibitem[Cronin and Weingart(2019)]%
        {cronin2019conflict}
\bibfield{author}{\bibinfo{person}{Matthew~A Cronin} {and} \bibinfo{person}{Laurie~R Weingart}.} \bibinfo{year}{2019}\natexlab{}.
\newblock \showarticletitle{Conflict across representational gaps: Threats to and opportunities for improved communication}.
\newblock \bibinfo{journal}{\emph{Proceedings of the National Academy of Sciences}} \bibinfo{volume}{116}, \bibinfo{number}{16} (\bibinfo{year}{2019}), \bibinfo{pages}{7642--7649}.
\newblock


\bibitem[Daily(2023)]%
        {GuangzhouDaily2023}
\bibfield{author}{\bibinfo{person}{Guangzhou Daily}.} \bibinfo{year}{2023}\natexlab{}.
\newblock \bibinfo{booktitle}{\emph{2023 Postdoctoral Researchers Recruitment Announcement, Guangzhou Daily Newspaper Group Postdoctoral Research Station}}.
\newblock
\urldef\tempurl%
\url{https://mp.weixin.qq.com/s/2sNgSZG8OChlvb5j3QSrtQ}
\showURL{%
\tempurl}


\bibitem[D'amour et~al\mbox{.}(2005)]%
        {d2005conceptual}
\bibfield{author}{\bibinfo{person}{Danielle D'amour}, \bibinfo{person}{Marcela Ferrada-Videla}, \bibinfo{person}{Leticia San Martin~Rodriguez}, {and} \bibinfo{person}{Marie-Dominique Beaulieu}.} \bibinfo{year}{2005}\natexlab{}.
\newblock \showarticletitle{The conceptual basis for interprofessional collaboration: Core concepts and theoretical frameworks}.
\newblock \bibinfo{journal}{\emph{Journal of interprofessional care}} \bibinfo{volume}{19}, \bibinfo{number}{sup1} (\bibinfo{year}{2005}), \bibinfo{pages}{116--131}.
\newblock


\bibitem[de~Assis(2024)]%
        {deAssis2024AI}
\bibfield{author}{\bibinfo{person}{Carolina de Assis}.} \bibinfo{year}{2024}\natexlab{}.
\newblock \bibinfo{title}{AI in Local Journalism: How Two Brazilian Media Outlets Are Using Generative AI to Provide Services to Their Audiences}.
\newblock
\newblock
\urldef\tempurl%
\url{https://latamjournalismreview.org/articles/ai-in-local-journalism-how-two-brazilian-media-outlets-are-using-generative-ai-to-provide-services-to-their-audiences/}
\showURL{%
\tempurl}
\newblock
\shownote{Accessed: 2024-09-12}.


\bibitem[de~Lima-Santos and Ceron(2021)]%
        {de2021artificial}
\bibfield{author}{\bibinfo{person}{Mathias-Felipe de Lima-Santos} {and} \bibinfo{person}{Wilson Ceron}.} \bibinfo{year}{2021}\natexlab{}.
\newblock \showarticletitle{Artificial intelligence in news media: current perceptions and future outlook}.
\newblock \bibinfo{journal}{\emph{Journalism and media}} \bibinfo{volume}{3}, \bibinfo{number}{1} (\bibinfo{year}{2021}), \bibinfo{pages}{13--26}.
\newblock


\bibitem[Deng et~al\mbox{.}(2023a)]%
        {deng2023understanding}
\bibfield{author}{\bibinfo{person}{Wesley~Hanwen Deng}, \bibinfo{person}{Boyuan Guo}, \bibinfo{person}{Alicia Devrio}, \bibinfo{person}{Hong Shen}, \bibinfo{person}{Motahhare Eslami}, {and} \bibinfo{person}{Kenneth Holstein}.} \bibinfo{year}{2023}\natexlab{a}.
\newblock \showarticletitle{Understanding Practices, Challenges, and Opportunities for User-Engaged Algorithm Auditing in Industry Practice}. In \bibinfo{booktitle}{\emph{Proceedings of the 2023 CHI Conference on Human Factors in Computing Systems}}. \bibinfo{pages}{1--18}.
\newblock


\bibitem[Deng et~al\mbox{.}(2023b)]%
        {deng2023investigating}
\bibfield{author}{\bibinfo{person}{Wesley~Hanwen Deng}, \bibinfo{person}{Nur Yildirim}, \bibinfo{person}{Monica Chang}, \bibinfo{person}{Motahhare Eslami}, \bibinfo{person}{Kenneth Holstein}, {and} \bibinfo{person}{Michael Madaio}.} \bibinfo{year}{2023}\natexlab{b}.
\newblock \showarticletitle{Investigating Practices and Opportunities for Cross-functional Collaboration around AI Fairness in Industry Practice}. In \bibinfo{booktitle}{\emph{Proceedings of the 2023 ACM Conference on Fairness, Accountability, and Transparency}}. \bibinfo{pages}{705--716}.
\newblock


\bibitem[Ding(2018)]%
        {ding2018deciphering}
\bibfield{author}{\bibinfo{person}{Jeffrey Ding}.} \bibinfo{year}{2018}\natexlab{}.
\newblock \showarticletitle{Deciphering China’s AI dream}.
\newblock \bibinfo{journal}{\emph{Future of Humanity Institute Technical Report}} (\bibinfo{year}{2018}).
\newblock


\bibitem[Dodds et~al\mbox{.}(2024a)]%
        {dodds2024collaborative}
\bibfield{author}{\bibinfo{person}{Tom{\'a}s Dodds}, \bibinfo{person}{Valeria Res{\'e}ndez}, \bibinfo{person}{Gerret von Nordheim}, \bibinfo{person}{Theo Araujo}, {and} \bibinfo{person}{Judith Moeller}.} \bibinfo{year}{2024}\natexlab{a}.
\newblock \showarticletitle{Collaborative Coding Cultures: How Journalists Use GitHub as a Trading Zone}.
\newblock \bibinfo{journal}{\emph{Digital Journalism}} (\bibinfo{year}{2024}), \bibinfo{pages}{1--22}.
\newblock


\bibitem[Dodds et~al\mbox{.}(2024b)]%
        {dodds2024impact}
\bibfield{author}{\bibinfo{person}{Tom{\'a}s Dodds}, \bibinfo{person}{Astrid Vandendaele}, \bibinfo{person}{Felix~M Simon}, \bibinfo{person}{Natali Helberger}, \bibinfo{person}{Valeria Resendez}, {and} \bibinfo{person}{Wang~Ngai Yeung}.} \bibinfo{year}{2024}\natexlab{b}.
\newblock \showarticletitle{The Impact of Knowledge Silos on Responsible AI Practices in Journalism}.
\newblock \bibinfo{journal}{\emph{arXiv preprint arXiv:2410.01138}} (\bibinfo{year}{2024}).
\newblock


\bibitem[Dove et~al\mbox{.}(2017)]%
        {dove2017ux}
\bibfield{author}{\bibinfo{person}{Graham Dove}, \bibinfo{person}{Kim Halskov}, \bibinfo{person}{Jodi Forlizzi}, {and} \bibinfo{person}{John Zimmerman}.} \bibinfo{year}{2017}\natexlab{}.
\newblock \showarticletitle{UX design innovation: Challenges for working with machine learning as a design material}. In \bibinfo{booktitle}{\emph{Proceedings of the 2017 chi conference on human factors in computing systems}}. \bibinfo{pages}{278--288}.
\newblock


\bibitem[Doyle et~al\mbox{.}(2016)]%
        {doyle2016viewing}
\bibfield{author}{\bibinfo{person}{P. Doyle}, \bibinfo{person}{M. Gelman}, {and} \bibinfo{person}{S. Gill}.} \bibinfo{year}{2016}\natexlab{}.
\newblock \bibinfo{booktitle}{\emph{Viewing the Future? Virtual Reality in Journalism}}.
\newblock \bibinfo{type}{Report}. \bibinfo{institution}{Knight Foundation}.
\newblock


\bibitem[Employment(2020)]%
        {PeoplePostdoc2020}
\bibfield{author}{\bibinfo{person}{Tsinghua Employment}.} \bibinfo{year}{2020}\natexlab{}.
\newblock \bibinfo{booktitle}{\emph{Recruitment | Postdoctoral Recruitment at People's Daily Online Postdoctoral Research Station}}.
\newblock
\urldef\tempurl%
\url{https://mp.weixin.qq.com/s/GlxD7rVt0P2wI1RJiHBS_A}
\showURL{%
\tempurl}


\bibitem[Fang et~al\mbox{.}(2024)]%
        {fang2024bias}
\bibfield{author}{\bibinfo{person}{Xiao Fang}, \bibinfo{person}{Shangkun Che}, \bibinfo{person}{Minjia Mao}, \bibinfo{person}{Hongzhe Zhang}, \bibinfo{person}{Ming Zhao}, {and} \bibinfo{person}{Xiaohang Zhao}.} \bibinfo{year}{2024}\natexlab{}.
\newblock \showarticletitle{Bias of AI-generated content: an examination of news produced by large language models}.
\newblock \bibinfo{journal}{\emph{Scientific Reports}} \bibinfo{volume}{14}, \bibinfo{number}{1} (\bibinfo{year}{2024}), \bibinfo{pages}{5224}.
\newblock


\bibitem[Fereday and Muir-Cochrane(2006)]%
        {fereday2006demonstrating}
\bibfield{author}{\bibinfo{person}{Jennifer Fereday} {and} \bibinfo{person}{Eimear Muir-Cochrane}.} \bibinfo{year}{2006}\natexlab{}.
\newblock \showarticletitle{Demonstrating rigor using thematic analysis: A hybrid approach of inductive and deductive coding and theme development}.
\newblock \bibinfo{journal}{\emph{International journal of qualitative methods}} \bibinfo{volume}{5}, \bibinfo{number}{1} (\bibinfo{year}{2006}), \bibinfo{pages}{80--92}.
\newblock


\bibitem[Fletcher and Nielsen(2024)]%
        {Fletcher2024}
\bibfield{author}{\bibinfo{person}{Richard Fletcher} {and} \bibinfo{person}{Rasmus~Kleis Nielsen}.} \bibinfo{year}{2024}\natexlab{}.
\newblock \showarticletitle{What does the public in six countries think of generative AI in news?}
\newblock \bibinfo{journal}{\emph{Reuters Institute for the Study of Journalism}} (\bibinfo{year}{2024}).
\newblock
\urldef\tempurl%
\url{https://doi.org/10.60625/RISJ-4ZB8-CG87}
\showURL{%
\tempurl}


\bibitem[Galison(2010)]%
        {galison2010trading}
\bibfield{author}{\bibinfo{person}{Peter Galison}.} \bibinfo{year}{2010}\natexlab{}.
\newblock \showarticletitle{Trading with the enemy}.
\newblock  (\bibinfo{year}{2010}).
\newblock


\bibitem[Goldstein(2017)]%
        {goldstein2017reflexivity}
\bibfield{author}{\bibinfo{person}{Shari~E Goldstein}.} \bibinfo{year}{2017}\natexlab{}.
\newblock \showarticletitle{Reflexivity in narrative research: Accessing meaning through the participant-researcher relationship.}
\newblock \bibinfo{journal}{\emph{Qualitative Psychology}} \bibinfo{volume}{4}, \bibinfo{number}{2} (\bibinfo{year}{2017}), \bibinfo{pages}{149}.
\newblock


\bibitem[Gutierrez~Lopez et~al\mbox{.}(2023)]%
        {gutierrez2023question}
\bibfield{author}{\bibinfo{person}{Marisela Gutierrez~Lopez}, \bibinfo{person}{Colin Porlezza}, \bibinfo{person}{Glenda Cooper}, \bibinfo{person}{Stephann Makri}, \bibinfo{person}{Andrew MacFarlane}, {and} \bibinfo{person}{Sondess Missaoui}.} \bibinfo{year}{2023}\natexlab{}.
\newblock \showarticletitle{A question of design: Strategies for embedding AI-driven tools into journalistic work routines}.
\newblock \bibinfo{journal}{\emph{Digital Journalism}} \bibinfo{volume}{11}, \bibinfo{number}{3} (\bibinfo{year}{2023}), \bibinfo{pages}{484--503}.
\newblock


\bibitem[Holbert et~al\mbox{.}(2020)]%
        {holbert2020afrofuturism}
\bibfield{author}{\bibinfo{person}{Nathan Holbert}, \bibinfo{person}{Mara Dando}, {and} \bibinfo{person}{Irani Correa}.} \bibinfo{year}{2020}\natexlab{}.
\newblock \showarticletitle{Afrofuturism as critical constructionist design: Building Futures from the past and present}.
\newblock \bibinfo{journal}{\emph{Learning, Media and Technology}} \bibinfo{volume}{45}, \bibinfo{number}{4} (\bibinfo{year}{2020}), \bibinfo{pages}{328--344}.
\newblock
\urldef\tempurl%
\url{https://doi.org/10.1080/17439884.2020.1754237}
\showDOI{\tempurl}


\bibitem[{International Center for Journalists (ICFJ)}(2017)]%
        {ICFJ2017}
\bibfield{author}{\bibinfo{person}{{International Center for Journalists (ICFJ)}}.} \bibinfo{year}{2017}\natexlab{}.
\newblock \bibinfo{title}{First-ever Global Survey of News Tech Reveals Perilous Digital Skills Gap}.
\newblock
\newblock
\urldef\tempurl%
\url{https://www.icfj.org/news/first-ever-global-survey-news-tech-reveals-perilous-digital-skills-gap}
\showURL{%
\tempurl}
\newblock
\shownote{Accessed: 2024-09-07}.


\bibitem[Jassawalla and Sashittal(1999)]%
        {jassawalla1999building}
\bibfield{author}{\bibinfo{person}{Avan~R Jassawalla} {and} \bibinfo{person}{Hemant~C Sashittal}.} \bibinfo{year}{1999}\natexlab{}.
\newblock \showarticletitle{Building collaborative cross-functional new product teams}.
\newblock \bibinfo{journal}{\emph{Academy of Management Perspectives}} \bibinfo{volume}{13}, \bibinfo{number}{3} (\bibinfo{year}{1999}), \bibinfo{pages}{50--63}.
\newblock


\bibitem[Jones(2024)]%
        {jones2024karen}
\bibfield{author}{\bibinfo{person}{Hessie Jones}.} \bibinfo{year}{2024}\natexlab{}.
\newblock \showarticletitle{Karen Hao Empowers Journalists To Unravel AI’s Complexities And Impacts}.
\newblock \bibinfo{journal}{\emph{Forbes}} (\bibinfo{date}{25 July} \bibinfo{year}{2024}).
\newblock
\urldef\tempurl%
\url{https://www.forbes.com/sites/hessiejones/2024/07/25/karen-hao-empowers-journalists-to-unravel-ais-complexities-and-impacts/}
\showURL{%
\tempurl}
\newblock
\shownote{Accessed: 2024-08-01}.


\bibitem[{JournalismAI}(2023)]%
        {journalismai_ai_2023}
\bibfield{author}{\bibinfo{person}{{JournalismAI}}.} \bibinfo{year}{2023}\natexlab{}.
\newblock \bibinfo{title}{How newsrooms around the world use AI: A JournalismAI 2023 global survey}.
\newblock
\newblock
\urldef\tempurl%
\url{https://blogs.lse.ac.uk/polis/2023/06/26/how-newsrooms-around-the-world-use-ai-a-journalismai-2023-global-survey/}
\showURL{%
\tempurl}
\newblock
\shownote{Accessed: 2024-07-22}.


\bibitem[Kahn(2024)]%
        {Kahn2024}
\bibfield{author}{\bibinfo{person}{Gretel Kahn}.} \bibinfo{year}{2024}\natexlab{}.
\newblock \bibinfo{title}{This newsroom has been experimenting with AI since 2020. Here is what they have learned}.
\newblock
\newblock
\urldef\tempurl%
\url{https://reutersinstitute.politics.ox.ac.uk/news/newsroom-has-been-experimenting-ai-2020-here-what-they-have-learned}
\showURL{%
\tempurl}
\newblock
\shownote{Accessed: 2024-09-10}.


\bibitem[Kasavin(2017)]%
        {kasavin2017trading}
\bibfield{author}{\bibinfo{person}{Ilya Kasavin}.} \bibinfo{year}{2017}\natexlab{}.
\newblock \showarticletitle{Trading zones as a subject-matter of social philosophy of science}.
\newblock \bibinfo{journal}{\emph{Epistemology \& Philosophy of Science}} \bibinfo{volume}{51}, \bibinfo{number}{1} (\bibinfo{year}{2017}), \bibinfo{pages}{8--17}.
\newblock


\bibitem[Kothari and Cruikshank(2022)]%
        {kothari2022artificial}
\bibfield{author}{\bibinfo{person}{Ammina Kothari} {and} \bibinfo{person}{Sally~Ann Cruikshank}.} \bibinfo{year}{2022}\natexlab{}.
\newblock \showarticletitle{Artificial intelligence and journalism: An Agenda for journalism research in Africa}.
\newblock \bibinfo{journal}{\emph{African Journalism Studies}} \bibinfo{volume}{43}, \bibinfo{number}{1} (\bibinfo{year}{2022}), \bibinfo{pages}{17--33}.
\newblock


\bibitem[Kuo(2024)]%
        {kuo2024china}
\bibfield{author}{\bibinfo{person}{Mercy~A. Kuo}.} \bibinfo{year}{2024}\natexlab{}.
\newblock \showarticletitle{China’s National Power and Artificial Intelligence: Insights from William C. Hannas}.
\newblock \bibinfo{journal}{\emph{The Diplomat}} (\bibinfo{year}{2024}).
\newblock
\urldef\tempurl%
\url{https://thediplomat.com/2024/07/chinas-national-power-and-artificial-intelligence/}
\showURL{%
\tempurl}


\bibitem[Kuo et~al\mbox{.}(2023)]%
        {kuo2023understanding}
\bibfield{author}{\bibinfo{person}{Tzu-Sheng Kuo}, \bibinfo{person}{Hong Shen}, \bibinfo{person}{Jisoo Geum}, \bibinfo{person}{Nev Jones}, \bibinfo{person}{Jason~I Hong}, \bibinfo{person}{Haiyi Zhu}, {and} \bibinfo{person}{Kenneth Holstein}.} \bibinfo{year}{2023}\natexlab{}.
\newblock \showarticletitle{Understanding Frontline Workers’ and Unhoused Individuals’ Perspectives on AI Used in Homeless Services}. In \bibinfo{booktitle}{\emph{Proceedings of the 2023 CHI Conference on Human Factors in Computing Systems}}. \bibinfo{pages}{1--17}.
\newblock


\bibitem[Laboratory(2023)]%
        {shlab2023}
\bibfield{author}{\bibinfo{person}{Shanghai~AI Laboratory}.} \bibinfo{year}{2023}\natexlab{}.
\newblock \bibinfo{title}{Shanghai AI Laboratory and China Media Group Jointly Release CCTV Listening Media Large Model (In Chinese)}.
\newblock
\newblock
\urldef\tempurl%
\url{https://www.shlab.org.cn/news/5443449}
\showURL{%
\tempurl}
\newblock
\shownote{Source: Shanghai Artificial Intelligence Laboratory}.


\bibitem[Lee et~al\mbox{.}(2020)]%
        {lee2020predicting}
\bibfield{author}{\bibinfo{person}{Sangwon Lee}, \bibinfo{person}{Seungahn Nah}, \bibinfo{person}{Deborah~S Chung}, {and} \bibinfo{person}{Junghwan Kim}.} \bibinfo{year}{2020}\natexlab{}.
\newblock \showarticletitle{Predicting AI news credibility: Communicative or social capital or both?}
\newblock In \bibinfo{booktitle}{\emph{Communicating Artificial Intelligence (AI)}}. \bibinfo{publisher}{Routledge}, \bibinfo{pages}{60--79}.
\newblock


\bibitem[Leiser(2022)]%
        {leiser2022bias}
\bibfield{author}{\bibinfo{person}{MR Leiser}.} \bibinfo{year}{2022}\natexlab{}.
\newblock \showarticletitle{Bias, journalistic endeavours, and the risks of artificial intelligence}.
\newblock In \bibinfo{booktitle}{\emph{Artificial Intelligence and the Media}}. \bibinfo{publisher}{Edward Elgar Publishing}, \bibinfo{pages}{8--32}.
\newblock


\bibitem[Lewis and Usher(2016)]%
        {lewis2016trading}
\bibfield{author}{\bibinfo{person}{Seth~C Lewis} {and} \bibinfo{person}{Nikki Usher}.} \bibinfo{year}{2016}\natexlab{}.
\newblock \showarticletitle{Trading zones, boundary objects, and the pursuit of news innovation: A case study of journalists and programmers}.
\newblock \bibinfo{journal}{\emph{Convergence}} \bibinfo{volume}{22}, \bibinfo{number}{5} (\bibinfo{year}{2016}), \bibinfo{pages}{543--560}.
\newblock


\bibitem[Li et~al\mbox{.}(2023)]%
        {li2023newsdialogues}
\bibfield{author}{\bibinfo{person}{Siheng Li}, \bibinfo{person}{Yichun Yin}, \bibinfo{person}{Cheng Yang}, \bibinfo{person}{Wangjie Jiang}, \bibinfo{person}{Yiwei Li}, \bibinfo{person}{Zesen Cheng}, \bibinfo{person}{Lifeng Shang}, \bibinfo{person}{Xin Jiang}, \bibinfo{person}{Qun Liu}, {and} \bibinfo{person}{Yujiu Yang}.} \bibinfo{year}{2023}\natexlab{}.
\newblock \showarticletitle{Newsdialogues: Towards proactive news grounded conversation}.
\newblock \bibinfo{journal}{\emph{arXiv preprint arXiv:2308.06501}} (\bibinfo{year}{2023}).
\newblock


\bibitem[Liu and Xu(2018)]%
        {liu2018}
\bibfield{author}{\bibinfo{person}{Xingchen Liu} {and} \bibinfo{person}{Qian Xu}.} \bibinfo{year}{2018}\natexlab{}.
\newblock \showarticletitle{Paradigm Impact and Value Reconstruction of Artificial Intelligence on Journalism and Publishing (In Chinese)}.
\newblock \bibinfo{journal}{\emph{Science and Publishing}} \bibinfo{number}{06} (\bibinfo{year}{2018}), \bibinfo{pages}{140--144}.
\newblock
\urldef\tempurl%
\url{https://doi.org/10.16510/j.cnki.kjycb.20180517.001}
\showDOI{\tempurl}


\bibitem[Maiden et~al\mbox{.}(2019)]%
        {maiden2019evaluating}
\bibfield{author}{\bibinfo{person}{Neil Maiden}, \bibinfo{person}{Konstantinos Zachos}, \bibinfo{person}{Amanda Brown}, \bibinfo{person}{Lars Nyre}, \bibinfo{person}{Balder Holm}, \bibinfo{person}{Aleksander~Nyg{\aa}rd Tonheim}, \bibinfo{person}{Claus Hesseling}, \bibinfo{person}{Andrea Wagemans}, {and} \bibinfo{person}{Dimitris Apostolou}.} \bibinfo{year}{2019}\natexlab{}.
\newblock \showarticletitle{Evaluating the use of digital creativity support by journalists in newsrooms}. In \bibinfo{booktitle}{\emph{Proceedings of the 2019 Conference on Creativity and Cognition}}. \bibinfo{pages}{222--232}.
\newblock


\bibitem[Majchrzak et~al\mbox{.}(2012)]%
        {majchrzak2012transcending}
\bibfield{author}{\bibinfo{person}{Ann Majchrzak}, \bibinfo{person}{Philip~HB More}, {and} \bibinfo{person}{Samer Faraj}.} \bibinfo{year}{2012}\natexlab{}.
\newblock \showarticletitle{Transcending knowledge differences in cross-functional teams}.
\newblock \bibinfo{journal}{\emph{Organization science}} \bibinfo{volume}{23}, \bibinfo{number}{4} (\bibinfo{year}{2012}), \bibinfo{pages}{951--970}.
\newblock


\bibitem[Min and Fink(2021)]%
        {min2021keeping}
\bibfield{author}{\bibinfo{person}{Seong~Jae Min} {and} \bibinfo{person}{Katherine Fink}.} \bibinfo{year}{2021}\natexlab{}.
\newblock \showarticletitle{Keeping up with the technologies: Distressed journalistic labor in the pursuit of “shiny” technologies}.
\newblock \bibinfo{journal}{\emph{Journalism Studies}} \bibinfo{volume}{22}, \bibinfo{number}{14} (\bibinfo{year}{2021}), \bibinfo{pages}{1987--2004}.
\newblock


\bibitem[M{\o}ller(2023)]%
        {moller2023bridging}
\bibfield{author}{\bibinfo{person}{Lynge~Asbj{\o}rn M{\o}ller}.} \bibinfo{year}{2023}\natexlab{}.
\newblock \showarticletitle{Bridging the tech-editorial gap: Lessons from two case studies of the development and integration of algorithmic curation in journalism}.
\newblock \bibinfo{journal}{\emph{Journalism Studies}} \bibinfo{volume}{24}, \bibinfo{number}{11} (\bibinfo{year}{2023}), \bibinfo{pages}{1458--1475}.
\newblock


\bibitem[Morgan and Nica(2020)]%
        {morgan2020iterative}
\bibfield{author}{\bibinfo{person}{David~L Morgan} {and} \bibinfo{person}{Andreea Nica}.} \bibinfo{year}{2020}\natexlab{}.
\newblock \showarticletitle{Iterative thematic inquiry: A new method for analyzing qualitative data}.
\newblock \bibinfo{journal}{\emph{International Journal of Qualitative Methods}}  \bibinfo{volume}{19} (\bibinfo{year}{2020}), \bibinfo{pages}{1609406920955118}.
\newblock


\bibitem[{National Key Laboratory of Cognitive Content Communication, People's Daily}(2023)]%
        {RenminRibao2023}
\bibfield{author}{\bibinfo{person}{{National Key Laboratory of Cognitive Content Communication, People's Daily}}.} \bibinfo{year}{2023}\natexlab{}.
\newblock \bibinfo{booktitle}{\emph{2023 Recruitment Announcement (Beijing Household Registration Solution) (In Chinese)}}.
\newblock
\urldef\tempurl%
\url{https://mp.weixin.qq.com/s/ywsSXQISC-3ZzZCCPH-w3A}
\showURL{%
\tempurl}
\newblock
\shownote{{Media Circle Recruitment}}.


\bibitem[Nguyen and Hekman(2022)]%
        {nguyen2022new}
\bibfield{author}{\bibinfo{person}{Dennis Nguyen} {and} \bibinfo{person}{Erik Hekman}.} \bibinfo{year}{2022}\natexlab{}.
\newblock \showarticletitle{A ‘new arms race’? Framing China and the USA in AI news reporting: A comparative analysis of the Washington Post and South China Morning Post}.
\newblock \bibinfo{journal}{\emph{Global Media and China}} \bibinfo{volume}{7}, \bibinfo{number}{1} (\bibinfo{year}{2022}), \bibinfo{pages}{58--77}.
\newblock


\bibitem[Nguyen et~al\mbox{.}(2018)]%
        {nguyen2018cross}
\bibfield{author}{\bibinfo{person}{Nguyen~Phong Nguyen}, \bibinfo{person}{Liem~Viet Ngo}, \bibinfo{person}{Tania Bucic}, {and} \bibinfo{person}{Nguyen~Dong Phong}.} \bibinfo{year}{2018}\natexlab{}.
\newblock \showarticletitle{Cross-functional knowledge sharing, coordination and firm performance: The role of cross-functional competition}.
\newblock \bibinfo{journal}{\emph{Industrial Marketing Management}}  \bibinfo{volume}{71} (\bibinfo{year}{2018}), \bibinfo{pages}{123--134}.
\newblock


\bibitem[Nicolini et~al\mbox{.}(2012)]%
        {nicolini2012understanding}
\bibfield{author}{\bibinfo{person}{Davide Nicolini}, \bibinfo{person}{Jeanne Mengis}, {and} \bibinfo{person}{Jacky Swan}.} \bibinfo{year}{2012}\natexlab{}.
\newblock \showarticletitle{Understanding the role of objects in cross-disciplinary collaboration}.
\newblock \bibinfo{journal}{\emph{Organization science}} \bibinfo{volume}{23}, \bibinfo{number}{3} (\bibinfo{year}{2012}), \bibinfo{pages}{612--629}.
\newblock


\bibitem[Nishal(2024)]%
        {nishal2024blueprints}
\bibfield{author}{\bibinfo{person}{S. Nishal}.} \bibinfo{year}{2024}\natexlab{}.
\newblock \bibinfo{title}{Blueprints for evaluating AI in journalism}.
\newblock
\newblock
\urldef\tempurl%
\url{https://generative-ai-newsroom.com/blueprints-for-evaluating-ai-in-journalism-e702c9e8c4f3}
\showURL{%
\tempurl}
\newblock
\shownote{Accessed: 2024-07-22}.


\bibitem[Nishal et~al\mbox{.}(2024)]%
        {nishal2024understanding}
\bibfield{author}{\bibinfo{person}{Sachita Nishal}, \bibinfo{person}{Jasmine Sinchai}, {and} \bibinfo{person}{Nicholas Diakopoulos}.} \bibinfo{year}{2024}\natexlab{}.
\newblock \showarticletitle{Understanding Practices around Computational News Discovery Tools in the Domain of Science Journalism}.
\newblock \bibinfo{journal}{\emph{Proceedings of the ACM on Human-Computer Interaction}} \bibinfo{volume}{8}, \bibinfo{number}{CSCW1} (\bibinfo{year}{2024}), \bibinfo{pages}{1--36}.
\newblock


\bibitem[Noy(2008)]%
        {noy2008sampling}
\bibfield{author}{\bibinfo{person}{Chaim Noy}.} \bibinfo{year}{2008}\natexlab{}.
\newblock \showarticletitle{Sampling knowledge: The hermeneutics of snowball sampling in qualitative research}.
\newblock \bibinfo{journal}{\emph{International Journal of social research methodology}} \bibinfo{volume}{11}, \bibinfo{number}{4} (\bibinfo{year}{2008}), \bibinfo{pages}{327--344}.
\newblock


\bibitem[of~China(2023)]%
        {cac2023}
\bibfield{author}{\bibinfo{person}{Cyberspace~Administration of China}.} \bibinfo{year}{2023}\natexlab{}.
\newblock \bibinfo{title}{Interim Measures for the Management of Generative Artificial Intelligence Services (In Chinese)}.
\newblock
\newblock
\urldef\tempurl%
\url{https://www.cac.gov.cn/2023-07/13/c_1690898327029107.htm}
\showURL{%
\tempurl}


\bibitem[of~Human~Resources and Security(2023)]%
        {peopledaily2023}
\bibfield{author}{\bibinfo{person}{Ministry of Human~Resources} {and} \bibinfo{person}{Social Security}.} \bibinfo{year}{2023}\natexlab{}.
\newblock \bibinfo{title}{People's Daily 2024 Annual Public Recruitment Announcement (In Chinese)}.
\newblock
\newblock
\urldef\tempurl%
\url{https://www.mohrss.gov.cn/SYrlzyhshbzb/fwyd/SYkaoshizhaopin/zyhgjjgsydwgkzp/zpgg/202312/t20231218_510704.html}
\showURL{%
\tempurl}
\newblock
\shownote{Source: Bureau of Personnel Management of Public Institutions}.


\bibitem[Parasie(2022)]%
        {parasie2022computing}
\bibfield{author}{\bibinfo{person}{Sylvain Parasie}.} \bibinfo{year}{2022}\natexlab{}.
\newblock \bibinfo{booktitle}{\emph{Computing the News: Data Journalism and the Search for Objectivity}}.
\newblock \bibinfo{publisher}{Columbia University Press}.
\newblock


\bibitem[Pinto and Barbosa(2024)]%
        {pinto2024artificial}
\bibfield{author}{\bibinfo{person}{Mois{\'e}s~Costa Pinto} {and} \bibinfo{person}{Suzana~Oliveira Barbosa}.} \bibinfo{year}{2024}\natexlab{}.
\newblock \showarticletitle{Artificial Intelligence (AI) in Brazilian Digital Journalism: Historical Context and Innovative Processes}.
\newblock \bibinfo{journal}{\emph{Journalism and Media}} \bibinfo{volume}{5}, \bibinfo{number}{1} (\bibinfo{year}{2024}), \bibinfo{pages}{325--341}.
\newblock


\bibitem[Popoola et~al\mbox{.}(2024)]%
        {popoola2024conceptualizing}
\bibfield{author}{\bibinfo{person}{Oladapo~Adeboye Popoola}, \bibinfo{person}{Henry~Ejiga Adama}, \bibinfo{person}{Chukwuekem~David Okeke}, {and} \bibinfo{person}{Abiodun~Emmanuel Akinoso}.} \bibinfo{year}{2024}\natexlab{}.
\newblock \showarticletitle{Conceptualizing agile development in digital transformations: Theoretical foundations and practical applications}.
\newblock \bibinfo{journal}{\emph{Engineering Science \& Technology Journal}} \bibinfo{volume}{5}, \bibinfo{number}{4} (\bibinfo{year}{2024}), \bibinfo{pages}{1524--1541}.
\newblock


\bibitem[{Public Media Alliance}(2024)]%
        {publicmedia2024}
\bibfield{author}{\bibinfo{person}{{Public Media Alliance}}.} \bibinfo{year}{2024}\natexlab{}.
\newblock \bibinfo{title}{Public Service Media and Generative AI}.
\newblock
\newblock
\urldef\tempurl%
\url{https://www.publicmediaalliance.org/resources/public-service-media-and-generative-ai/}
\showURL{%
\tempurl}
\newblock
\shownote{Accessed: 2024-09-10}.


\bibitem[Sadek et~al\mbox{.}(2023)]%
        {sadek2023designing}
\bibfield{author}{\bibinfo{person}{Malak Sadek}, \bibinfo{person}{Rafael~A Calvo}, {and} \bibinfo{person}{C{\'e}line Mougenot}.} \bibinfo{year}{2023}\natexlab{}.
\newblock \showarticletitle{Designing value-sensitive AI: a critical review and recommendations for socio-technical design processes}.
\newblock \bibinfo{journal}{\emph{AI and Ethics}} (\bibinfo{year}{2023}), \bibinfo{pages}{1--19}.
\newblock


\bibitem[Salvato et~al\mbox{.}(2017)]%
        {salvato2017cooperation}
\bibfield{author}{\bibinfo{person}{Carlo Salvato}, \bibinfo{person}{Jeffrey~J Reuer}, {and} \bibinfo{person}{Pierpaolo Battigalli}.} \bibinfo{year}{2017}\natexlab{}.
\newblock \showarticletitle{Cooperation across disciplines: A multilevel perspective on cooperative behavior in governing interfirm relations}.
\newblock \bibinfo{journal}{\emph{Academy of Management Annals}} \bibinfo{volume}{11}, \bibinfo{number}{2} (\bibinfo{year}{2017}), \bibinfo{pages}{960--1004}.
\newblock


\bibitem[Seychell et~al\mbox{.}(2024)]%
        {seychell2024ai}
\bibfield{author}{\bibinfo{person}{Dylan Seychell}, \bibinfo{person}{Gabriel Hili}, \bibinfo{person}{Jonathan Attard}, {and} \bibinfo{person}{Konstaninos Makantatis}.} \bibinfo{year}{2024}\natexlab{}.
\newblock \showarticletitle{AI as a Tool for Fair Journalism: Case Studies from Malta}. In \bibinfo{booktitle}{\emph{2024 IEEE Conference on Artificial Intelligence (CAI)}}. IEEE, \bibinfo{pages}{127--132}.
\newblock


\bibitem[Shen and Chen(2018)]%
        {shen2018}
\bibfield{author}{\bibinfo{person}{Nan Shen} {and} \bibinfo{person}{Yihua Chen}.} \bibinfo{year}{2018}\natexlab{}.
\newblock \showarticletitle{Exploring the "MAGIC" of Intelligent News Production (In Chinese)}.
\newblock \bibinfo{journal}{\emph{News Frontline}} \bibinfo{number}{15} (\bibinfo{year}{2018}), \bibinfo{pages}{75--77}.
\newblock


\bibitem[Shrager(2010)]%
        {shrager2010wizards}
\bibfield{author}{\bibinfo{person}{Jeff Shrager}.} \bibinfo{year}{2010}\natexlab{}.
\newblock \showarticletitle{From wizards to trading zones: Crossing the chasm of computers in scientific collaboration}.
\newblock  (\bibinfo{year}{2010}).
\newblock


\bibitem[Shuhang(2023)]%
        {shu2023}
\bibfield{author}{\bibinfo{person}{Shuhang}.} \bibinfo{year}{2023}\natexlab{}.
\newblock \bibinfo{title}{Television Stations Going Crazy Over AIGC with an Official Announcement Every Two Days on Average (In Chinese)}.
\newblock
\newblock
\urldef\tempurl%
\url{https://www.163.com/dy/article/IV9R1QA605118BEE.html}
\showURL{%
\tempurl}


\bibitem[Simmons(2022)]%
        {simmons2022political}
\bibfield{author}{\bibinfo{person}{Omari~Scott Simmons}.} \bibinfo{year}{2022}\natexlab{}.
\newblock \showarticletitle{Political Risk Management}.
\newblock \bibinfo{journal}{\emph{Wm. \& Mary L. Rev.}}  \bibinfo{volume}{64} (\bibinfo{year}{2022}), \bibinfo{pages}{707}.
\newblock


\bibitem[Simon(2022)]%
        {simon2022uneasy}
\bibfield{author}{\bibinfo{person}{Felix~M Simon}.} \bibinfo{year}{2022}\natexlab{}.
\newblock \showarticletitle{Uneasy bedfellows: AI in the news, platform companies and the issue of journalistic autonomy}.
\newblock \bibinfo{journal}{\emph{Digital journalism}} \bibinfo{volume}{10}, \bibinfo{number}{10} (\bibinfo{year}{2022}), \bibinfo{pages}{1832--1854}.
\newblock


\bibitem[Simon(2024a)]%
        {simon_ai_2024}
\bibfield{author}{\bibinfo{person}{Felix~M Simon}.} \bibinfo{year}{2024}\natexlab{a}.
\newblock \showarticletitle{Artificial Intelligence in the news: How Ai Retools, rationalizes, and reshapes journalism and the Public Arena}.
\newblock \bibinfo{journal}{\emph{Columbia Journalism Review}} (\bibinfo{date}{6 February} \bibinfo{year}{2024}).
\newblock
\urldef\tempurl%
\url{https://www.cjr.org/tow_center_reports/artificial-intelligence-in-the-news.php}
\showURL{%
\tempurl}
\newblock
\shownote{Accessed: 2024-07-22}.


\bibitem[Simon(2024b)]%
        {simon2024escape}
\bibfield{author}{\bibinfo{person}{Felix~M Simon}.} \bibinfo{year}{2024}\natexlab{b}.
\newblock \showarticletitle{Escape me if you can: How ai reshapes news organisations’ dependency on platform companies}.
\newblock \bibinfo{journal}{\emph{Digital Journalism}} \bibinfo{volume}{12}, \bibinfo{number}{2} (\bibinfo{year}{2024}), \bibinfo{pages}{149--170}.
\newblock


\bibitem[Stenbom et~al\mbox{.}(2023)]%
        {stenbom2023exploring}
\bibfield{author}{\bibinfo{person}{Agnes Stenbom}, \bibinfo{person}{Mattias Wiggberg}, {and} \bibinfo{person}{Tobias Norlund}.} \bibinfo{year}{2023}\natexlab{}.
\newblock \showarticletitle{Exploring communicative AI: Reflections from a Swedish newsroom}.
\newblock \bibinfo{journal}{\emph{Digital Journalism}} \bibinfo{volume}{11}, \bibinfo{number}{9} (\bibinfo{year}{2023}), \bibinfo{pages}{1622--1640}.
\newblock


\bibitem[Stone et~al\mbox{.}(2016)]%
        {stone2016embedding}
\bibfield{author}{\bibinfo{person}{Maria Stone}, \bibinfo{person}{Frank Bentley}, \bibinfo{person}{Brooke White}, {and} \bibinfo{person}{Mike Shebanek}.} \bibinfo{year}{2016}\natexlab{}.
\newblock \showarticletitle{Embedding user understanding in the corporate culture: Ux research and accessibility at yahoo}. In \bibinfo{booktitle}{\emph{Proceedings of the 2016 CHI Conference Extended Abstracts on Human Factors in Computing Systems}}. \bibinfo{pages}{823--832}.
\newblock


\bibitem[Tahaei et~al\mbox{.}(2021)]%
        {tahaei2021privacy}
\bibfield{author}{\bibinfo{person}{Mohammad Tahaei}, \bibinfo{person}{Alisa Frik}, {and} \bibinfo{person}{Kami Vaniea}.} \bibinfo{year}{2021}\natexlab{}.
\newblock \showarticletitle{Privacy champions in software teams: Understanding their motivations, strategies, and challenges}. In \bibinfo{booktitle}{\emph{Proceedings of the 2021 CHI Conference on Human Factors in Computing Systems}}. \bibinfo{pages}{1--15}.
\newblock


\bibitem[{Tencent News}(2024)]%
        {tencent_ai_news_2024}
\bibfield{author}{\bibinfo{person}{{Tencent News}}.} \bibinfo{year}{2024}\natexlab{}.
\newblock \bibinfo{title}{CCTV bears the brunt of pioneering the use of AI for news footage, and the netizens aren’t buying it?}
\newblock
\newblock
\urldef\tempurl%
\url{https://new.qq.com/rain/a/20240412A06Z2W00}
\showURL{%
\tempurl}
\newblock
\shownote{Accessed: 2024-07-22}.


\bibitem[Thrasher(2024)]%
        {Thrasher2024}
\bibfield{author}{\bibinfo{person}{Michael Thrasher}.} \bibinfo{year}{2024}\natexlab{}.
\newblock \bibinfo{title}{Bloomberg’s First Generative AI Tool Hits the Terminal}.
\newblock
\newblock
\urldef\tempurl%
\url{https://www.institutionalinvestor.com/article/2cqjgsulkx3md4n3ox2ps/portfolio/bloombergs-first-generative-ai-tool-hits-the-terminal}
\showURL{%
\tempurl}
\newblock
\shownote{Accessed: 2024-09-10}.


\bibitem[Thurman et~al\mbox{.}(2017)]%
        {thurman2017reporters}
\bibfield{author}{\bibinfo{person}{Neil Thurman}, \bibinfo{person}{Konstantin D{\"o}rr}, {and} \bibinfo{person}{Jessica Kunert}.} \bibinfo{year}{2017}\natexlab{}.
\newblock \showarticletitle{When reporters get hands-on with robo-writing: Professionals consider automated journalism’s capabilities and consequences}.
\newblock \bibinfo{journal}{\emph{Digital journalism}} \bibinfo{volume}{5}, \bibinfo{number}{10} (\bibinfo{year}{2017}), \bibinfo{pages}{1240--1259}.
\newblock


\bibitem[Trattner et~al\mbox{.}(2022)]%
        {trattner2022responsible}
\bibfield{author}{\bibinfo{person}{Christoph Trattner}, \bibinfo{person}{Dietmar Jannach}, \bibinfo{person}{Enrico Motta}, \bibinfo{person}{Irene Costera~Meijer}, \bibinfo{person}{Nicholas Diakopoulos}, \bibinfo{person}{Mehdi Elahi}, \bibinfo{person}{Andreas~L Opdahl}, \bibinfo{person}{Bj{\o}rnar Tessem}, \bibinfo{person}{Nj{\aa}l Borch}, \bibinfo{person}{Morten Fjeld}, {et~al\mbox{.}}} \bibinfo{year}{2022}\natexlab{}.
\newblock \showarticletitle{Responsible media technology and AI: challenges and research directions}.
\newblock \bibinfo{journal}{\emph{AI and Ethics}} \bibinfo{volume}{2}, \bibinfo{number}{4} (\bibinfo{year}{2022}), \bibinfo{pages}{585--594}.
\newblock


\bibitem[Wagemans and Witschge(2019)]%
        {wagemans2019examining}
\bibfield{author}{\bibinfo{person}{Andrea Wagemans} {and} \bibinfo{person}{Tamara Witschge}.} \bibinfo{year}{2019}\natexlab{}.
\newblock \showarticletitle{Examining innovation as process: Action research in journalism studies}.
\newblock \bibinfo{journal}{\emph{Convergence}} \bibinfo{volume}{25}, \bibinfo{number}{2} (\bibinfo{year}{2019}), \bibinfo{pages}{209--224}.
\newblock


\bibitem[Walters(2024)]%
        {walters2024new}
\bibfield{author}{\bibinfo{person}{Patrick Walters}.} \bibinfo{year}{2024}\natexlab{}.
\newblock \showarticletitle{New Guests Crashing the Party: A Typology of Journalistic Collaboration}.
\newblock \bibinfo{journal}{\emph{Journalism Studies}} \bibinfo{volume}{25}, \bibinfo{number}{2} (\bibinfo{year}{2024}), \bibinfo{pages}{140--159}.
\newblock


\bibitem[W{\"o}lker and Powell(2021)]%
        {wolker2021algorithms}
\bibfield{author}{\bibinfo{person}{Anja W{\"o}lker} {and} \bibinfo{person}{Thomas~E Powell}.} \bibinfo{year}{2021}\natexlab{}.
\newblock \showarticletitle{Algorithms in the newsroom? News readers’ perceived credibility and selection of automated journalism}.
\newblock \bibinfo{journal}{\emph{Journalism}} \bibinfo{volume}{22}, \bibinfo{number}{1} (\bibinfo{year}{2021}), \bibinfo{pages}{86--103}.
\newblock


\bibitem[Yang et~al\mbox{.}(2018a)]%
        {yang2018mapping}
\bibfield{author}{\bibinfo{person}{Qian Yang}, \bibinfo{person}{Nikola Banovic}, {and} \bibinfo{person}{John Zimmerman}.} \bibinfo{year}{2018}\natexlab{a}.
\newblock \showarticletitle{Mapping machine learning advances from hci research to reveal starting places for design innovation}. In \bibinfo{booktitle}{\emph{Proceedings of the 2018 CHI conference on human factors in computing systems}}. \bibinfo{pages}{1--11}.
\newblock


\bibitem[Yang et~al\mbox{.}(2019)]%
        {yang2019sketching}
\bibfield{author}{\bibinfo{person}{Qian Yang}, \bibinfo{person}{Justin Cranshaw}, \bibinfo{person}{Saleema Amershi}, \bibinfo{person}{Shamsi~T. Iqbal}, {and} \bibinfo{person}{Jaime Teevan}.} \bibinfo{year}{2019}\natexlab{}.
\newblock \showarticletitle{Sketching NLP: A Case Study of Exploring the Right Things To Design with Language Intelligence}. In \bibinfo{booktitle}{\emph{Proceedings of the 2019 CHI Conference on Human Factors in Computing Systems}}. \bibinfo{publisher}{Association for Computing Machinery}, \bibinfo{address}{New York, NY, USA}.
\newblock
\urldef\tempurl%
\url{https://doi.org/10.1145/3290605.3300415}
\showDOI{\tempurl}


\bibitem[Yang et~al\mbox{.}(2018b)]%
        {yang2018investigating}
\bibfield{author}{\bibinfo{person}{Qian Yang}, \bibinfo{person}{Alex Scuito}, \bibinfo{person}{John Zimmerman}, \bibinfo{person}{Jodi Forlizzi}, {and} \bibinfo{person}{Aaron Steinfeld}.} \bibinfo{year}{2018}\natexlab{b}.
\newblock \showarticletitle{Investigating how experienced UX designers effectively work with machine learning}. In \bibinfo{booktitle}{\emph{Proceedings of the 2018 designing interactive systems conference}}. \bibinfo{pages}{585--596}.
\newblock


\bibitem[Yang et~al\mbox{.}(2020)]%
        {yang2020re}
\bibfield{author}{\bibinfo{person}{Qian Yang}, \bibinfo{person}{Aaron Steinfeld}, \bibinfo{person}{Carolyn Ros{\'e}}, {and} \bibinfo{person}{John Zimmerman}.} \bibinfo{year}{2020}\natexlab{}.
\newblock \showarticletitle{Re-examining whether, why, and how human-AI interaction is uniquely difficult to design}. In \bibinfo{booktitle}{\emph{Proceedings of the 2020 chi conference on human factors in computing systems}}. \bibinfo{pages}{1--13}.
\newblock


\bibitem[Yang et~al\mbox{.}(2018c)]%
        {yang2018grounding}
\bibfield{author}{\bibinfo{person}{Qian Yang}, \bibinfo{person}{Jungwoo Suh}, \bibinfo{person}{Nicholas Chen}, {and} \bibinfo{person}{Gonzalo Ramos}.} \bibinfo{year}{2018}\natexlab{c}.
\newblock \showarticletitle{Grounding Interactive Machine Learning Tool Design in how Non-Experts actually build models}. In \bibinfo{booktitle}{\emph{Proceedings of the 2018 Designing Interactive Systems Conference}}. \bibinfo{publisher}{Association for Computing Machinery}, \bibinfo{address}{New York, NY, USA}, \bibinfo{pages}{573--584}.
\newblock
\urldef\tempurl%
\url{https://doi.org/10.1145/3196709.3196729}
\showDOI{\tempurl}


\bibitem[Yao et~al\mbox{.}(2024)]%
        {yao2024exploring}
\bibfield{author}{\bibinfo{person}{Bingsheng Yao}, \bibinfo{person}{Yao Du}, \bibinfo{person}{Yue Fu}, \bibinfo{person}{Xuhai Xu}, \bibinfo{person}{Yanjun Gao}, \bibinfo{person}{Hong Yu}, {and} \bibinfo{person}{Dakuo Wang}.} \bibinfo{year}{2024}\natexlab{}.
\newblock \showarticletitle{Exploring Interdisciplinary Team Collaboration in Clinical NLP Projects Through the Lens of Activity Theory}.
\newblock \bibinfo{journal}{\emph{arXiv preprint arXiv:2410.00174}} (\bibinfo{year}{2024}).
\newblock


\bibitem[Yu et~al\mbox{.}(2017)]%
        {yu2017}
\bibfield{author}{\bibinfo{person}{Guoming Yu}, \bibinfo{person}{Meina Lan}, {and} \bibinfo{person}{Wei Li}.} \bibinfo{year}{2017}\natexlab{}.
\newblock \showarticletitle{Intelligentization: The Core Logic of Future Communication Mode Innovation—Also on the Basic Operation Paradigm of "Artificial Intelligence + Media" (In Chinese)}.
\newblock \bibinfo{journal}{\emph{Journalism and Writing}} \bibinfo{number}{03} (\bibinfo{year}{2017}), \bibinfo{pages}{41--45}.
\newblock


\bibitem[Zamith(2019)]%
        {zamith2019algorithms}
\bibfield{author}{\bibinfo{person}{Rodrigo Zamith}.} \bibinfo{year}{2019}\natexlab{}.
\newblock \showarticletitle{Algorithms and journalism}.
\newblock In \bibinfo{booktitle}{\emph{Oxford research encyclopedia of communication}}.
\newblock


\bibitem[Zeng(2020)]%
        {zeng2020artificial}
\bibfield{author}{\bibinfo{person}{Jinghan Zeng}.} \bibinfo{year}{2020}\natexlab{}.
\newblock \showarticletitle{Artificial intelligence and China's authoritarian governance}.
\newblock \bibinfo{journal}{\emph{International Affairs}} \bibinfo{volume}{96}, \bibinfo{number}{6} (\bibinfo{year}{2020}), \bibinfo{pages}{1441--1459}.
\newblock


\bibitem[Zhang et~al\mbox{.}(2020)]%
        {zhang2020data}
\bibfield{author}{\bibinfo{person}{Amy~X Zhang}, \bibinfo{person}{Michael Muller}, {and} \bibinfo{person}{Dakuo Wang}.} \bibinfo{year}{2020}\natexlab{}.
\newblock \showarticletitle{How do data science workers collaborate? roles, workflows, and tools}.
\newblock \bibinfo{journal}{\emph{Proceedings of the ACM on Human-Computer Interaction}} \bibinfo{volume}{4}, \bibinfo{number}{CSCW1} (\bibinfo{year}{2020}), \bibinfo{pages}{1--23}.
\newblock


\bibitem[Zhao and Ren(2021)]%
        {zhao2021}
\bibfield{author}{\bibinfo{person}{Ximin Zhao} {and} \bibinfo{person}{Zhiming Ren}.} \bibinfo{year}{2021}\natexlab{}.
\newblock \showarticletitle{Challenges and Value Reconstruction of Journalists' Competence in the Era of Artificial Intelligence (In Chinese)}.
\newblock \bibinfo{journal}{\emph{Media}} \bibinfo{number}{05} (\bibinfo{year}{2021}), \bibinfo{pages}{41--43}.
\newblock


\bibitem[Zhong and Zhang(2019)]%
        {zhong2019}
\bibfield{author}{\bibinfo{person}{Yingjiong Zhong} {and} \bibinfo{person}{Han Zhang}.} \bibinfo{year}{2019}\natexlab{}.
\newblock \showarticletitle{Kuai Bi Xiao Xin: Xinhua News Agency's First Robot Reporter}.
\newblock \bibinfo{journal}{\emph{People's Daily - Journalism Frontline}} (\bibinfo{date}{27 February} \bibinfo{year}{2019}).
\newblock
\urldef\tempurl%
\url{http://media.people.com.cn/GB/n1/2019/0227/c425664-30905230.html}
\showURL{%
\tempurl}


\end{thebibliography}

\appendix
\section{Appendix: Table ~\ref{tab:FormativeParticipant}. Participants’ Demographics}
\add{For Table ~\ref{tab:FormativeParticipant}, it should also be noted that although our demographic table provides a detailed categorization of cross-functional collaboration activities in which participants were involved, this classification primarily emerged from the subsequent analysis of the researchers. During the recruitment phase, we did not have such a comprehensive classification in place. Instead, we relied on the broader criteria we mentioned above to identify potential participants who had experience working in cross-functional settings. These broader criteria allowed us to capture a diverse range of collaborative experiences, which were later systematically categorized through our research process.}

\begin{table*}[ht]
  \centering
  \caption{~\add{Participants’ Demographics. Experience (yrs) refers to the total years of working experience in the news industry. The numbers in the involvement column correspond to different cross-functional collaboration activities:
1. Identifying the Collaboration Context (e.g., sharing goals in workshops);
2. Defining the Specific Task (e.g., discussing data requirements);
3. Designing the Tool (e.g., developing initial AI prototypes or conducting computational analysis);
4. Applying the Tool in the Newsroom (e.g., using automated tools in reporting);
5. Reflections on Cross-Functional Collaboration (e.g., team retrospectives);
6. Providing Data Annotations for Collaboration (e.g., supplying training data).
The two-letter codes represent distinct news organizations (NN, IA, RP, LM, ON). \textit{Noted}: Activities 1–5 are derived from our studies on journalists and AI technologists, aligning with ~\autoref{practices} of this paper. Activity 6, however, originates from our research on AI workers and primarily corresponds to \autoref{invisible_labour}.}}
  \label{tab:FormativeParticipant}
  \setlength{\tabcolsep}{10pt}
  \begin{tabularx}{\textwidth}{cccccccc}
    \toprule
    \textbf{ID} & \textbf{Gender} & \textbf{Age (yrs)} & \textbf{Experience (yrs)} & \textbf{Position} & \textbf{Education} & \textbf{Affiliation} & \textbf{Involvement} \\
    \midrule
    J1  & Female & 46-50 & 21-25 & Journalist       & Master   & NN & 1, 2, 3, 4, 5 \\
    J2  & Female & 41-45 & 16-20 & Journalist       & Bachelor & IA & 1, 2, 4, 5 \\
    J3  & Male   & 41-45 & 21-25 & Journalist       & Master   & RP & 1, 2, 3, 4 \\
    J4  & Female & 36-40 & 11-15 & Journalist       & Master   & LM & 1, 2, 4, 5, 6 \\
    J5  & Female & 31-35 & 11-15 & Journalist       & Master   & ON & 2, 3, 4, 5 \\
    J6  & Female & 31-35 & 6-10  & Journalist       & Bachelor & NN & 1, 2, 3, 4, 5 \\
    J7  & Male   & 36-40 & 6-10  & Journalist       & Master   & IA & 1, 2, 3, 4, 5 \\
    J8  & Male   & 31-35 & 11-15 & Journalist       & Master   & RP & 2, 3, 4, 5 \\
    J9  & Male   & 26-30 & 1-5   & Journalist       & Master   & LM & 1, 2, 4, 6 \\
    J10 & Male   & 26-30 & 1-5   & Journalist       & Master   & ON & 1, 2, 4, 5, 6 \\
    J11 & Female & 26-30 & 1-5   & Journalist       & Bachelor & NN & 1, 2, 3, 6 \\
    J12 & Female & 21-25 & 1-5   & Journalist       & Bachelor & IA & 1, 2, 4, 5, 6 \\
    J13 & Male   & 31-35 & 6-10  & Journalist       & Bachelor & RP & 1, 4, 5, 6 \\
    J14 & Female & 26-30 & 6-10  & Journalist       & Bachelor & LM & 1, 2, 3, 4, 5, 6 \\
    J15 & Female & 26-30 & 1-5   & Journalist       & Master   & ON & 1, 2, 3, 4, 5 \\
    J16 & Female & 21-25 & 1-5   & Journalist       & Bachelor & NN & 1, 2, 3, 4, 6 \\
    J17 & Male   & 21-25 & 1-5   & Journalist       & Bachelor & IA & 1, 2, 4, 5, 6 \\
    \midrule
    T1  & Male   & 26-30 & 1-5   & AI Technologist & PhD      & RP & 1, 2, 3, 4, 5 \\
    T2  & Male   & 26-30 & 1-5   & AI Technologist & PhD      & ON & 1, 3, 4, 5, 6 \\
    T3  & Male   & 26-30 & 1-5   & AI Technologist & Master   & NN & 2, 3, 4, 5 \\
    T4  & Male   & 26-30 & 1-5   & AI Technologist & Master   & IA & 1, 2, 3, 5 \\
    T5  & Male   & 26-30 & 1-5   & AI Technologist & Master   & LM & 1, 2, 3, 4, 5, 6 \\
    T6  & Male   & 26-30 & 1-5   & AI Technologist & Master   & RP & 1, 3, 4, 5, 6 \\
    \midrule
    W1  & Female & 21-25 & 1-5   & AI Worker       & Bachelor & IA & 6 \\
    W2  & Female & 21-25 & 1-5   & AI Worker       & Bachelor & LM & 6 \\
    W3  & Female & 21-25 & 1-5   & AI Worker       & Bachelor & ON & 6 \\
    \bottomrule
  \end{tabularx}
\end{table*}

\end{document}